\RequirePackage{fix-cm}
\documentclass[smallextended]{svjour3}   
\smartqed  

\usepackage{amsmath,amssymb,amsfonts}
\usepackage{graphicx}
\usepackage{subcaption}
\usepackage{xcolor}
\usepackage{tabularx}
\usepackage{xspace}
\usepackage[inline]{enumitem} 
\usepackage{booktabs}
\usepackage{multirow}
\usepackage{hyperref}
\graphicspath{{}}
\usepackage{cite}
\usepackage{color, colortbl}
\usepackage{caption}
\usepackage{tcolorbox}

\newtcolorbox[auto counter]{mybox}[1][]{%
    title=Finding~\thetcbcounter, arc=1pt,outer arc=1pt, #1}

\newcommand{\revision}[1]{#1}

\newcommand{\rev}[1]{#1}

\usepackage{cite}

\newcommand{\yake}{\texttt{YAKE!}\xspace}
\newcommand{\README}{\texttt{README}\xspace}
\newcommand{\lbl}[1]{`\textit{#1}'\xspace}
\newcommand*{\ie}{i.e.,\@\xspace}
\newcommand*{\eg}{e.g.,\@\xspace}

\journalname{Empirical Software Engineering}

\begin{document}

\title{Multi-granular Software Annotation using\\File-level Weak Labelling}



\titlerunning{Multi-granular Software Annotation using File-Level Weak Labelling}        

\author{Cezar Sas         \and
        Andrea Capiluppi 
}


\institute{Cezar Sas \at
            Bernoulli Institute, University of Groningen, Groningen, The Netherlands \\
            \email{c.a.sas@rug.nl}            \\
            ORCID: 0000-0002-3018-0140
           \and
           Andrea Capiluppi \at
        Bernoulli Institute, University of Groningen, Groningen, The Netherlands \\
            \email{a.capiluppi@rug.nl}   \\
            ORCID: 0000-0001-9469-6050
}

\date{Received: date / Accepted: date}

\maketitle

\begin{abstract}

One of the most time-consuming tasks for developers is the comprehension of new code bases. An effective approach to aid this process is to label source code files with meaningful annotations, which can help developers understand the content and functionality of a code base quicker. However, most existing solutions for code annotation focus on project-level classification: manually labelling individual files is time-consuming, error-prone and hard to scale.

The work presented in this paper aims to automate the annotation of files by leveraging project-level labels; and using the file-level annotations to annotate items at larger levels of granularity, for example, packages and a whole project. 

We propose a novel approach to annotate source code files using a weak labelling approach and a subsequent hierarchical aggregation. We investigate whether this approach is effective in achieving multi-granular annotations of software projects, which can aid developers in understanding the content and functionalities of a code base more quickly.

Our evaluation uses a combination of human assessment and automated metrics to evaluate the annotations' quality. Our approach correctly annotated 50\% of files and more than 50\% of packages. Moreover, the information captured at the file-level allowed us to identify, on average, three new relevant labels for any given project. 

We can conclude that the proposed approach is a convenient and promising way to generate noisy (not precise) annotations for files. Furthermore, hierarchical aggregation effectively preserves the information captured at file-level, and it can be propagated to packages and the overall project itself.

\end{abstract}

\section{Introduction}

Large code bases are becoming more common, both open-source and private. This rapid increase in software development translates into many developers switching to new projects, which requires considerable time to familiarize themselves with their content~\cite{xia2019measuring}.

In past research, several approaches have been proposed for automatic software application domain classification~\cite{izadi2020topic, nguyen2020automated, rocco2020topfilter, sipio2020naive}. Nevertheless, while showing promising results, past and current works have so far focused on classifying the project as a whole. Moreover, these approaches do not consider the compositionality of software, since a large system typically comprises several modules and components, each with its own functionality. As a result, several past and current approaches have only assigned a single label~\cite{sas2022antipatterns} to projects, and many rely on proxies like the \README file to infer labels.

Although there are instances of prior work focusing on the use of source code identifiers to assign topics to files~\cite{kuhn2007semantic}, they are based on the clustering of files with shared terms using Latent Semantic Analysis (LSA). This approach can be effective for a single project, but it still requires manual annotations of the clusters. Therefore, this solution is not scalable to large code bases, as it still requires substantive human intervention, in the form of manual annotations. When such annotations are unavailable, developers will still be required to understand the cluster, which can lead to ambiguity between developers due to the vagueness of natural language. %

Performing manual annotation of files is time-consuming and expensive, and as such, it requires automated methods to annotate data. However, weak supervision is a rapidly developing field in machine learning (ML), and it is a research area that has so far shown interesting results~\cite{zhang2022weak}: as a method, it focuses on training ML models using imprecise, incomplete, or noisy labels. The labels are created through weak labelling, an automatic approach to data annotation based on heuristics. 

\begin{figure}[htb!]
    \centering
    \begin{subfigure}[b]{0.45\textwidth}
         \centering
         \includegraphics[width=0.95\textwidth]{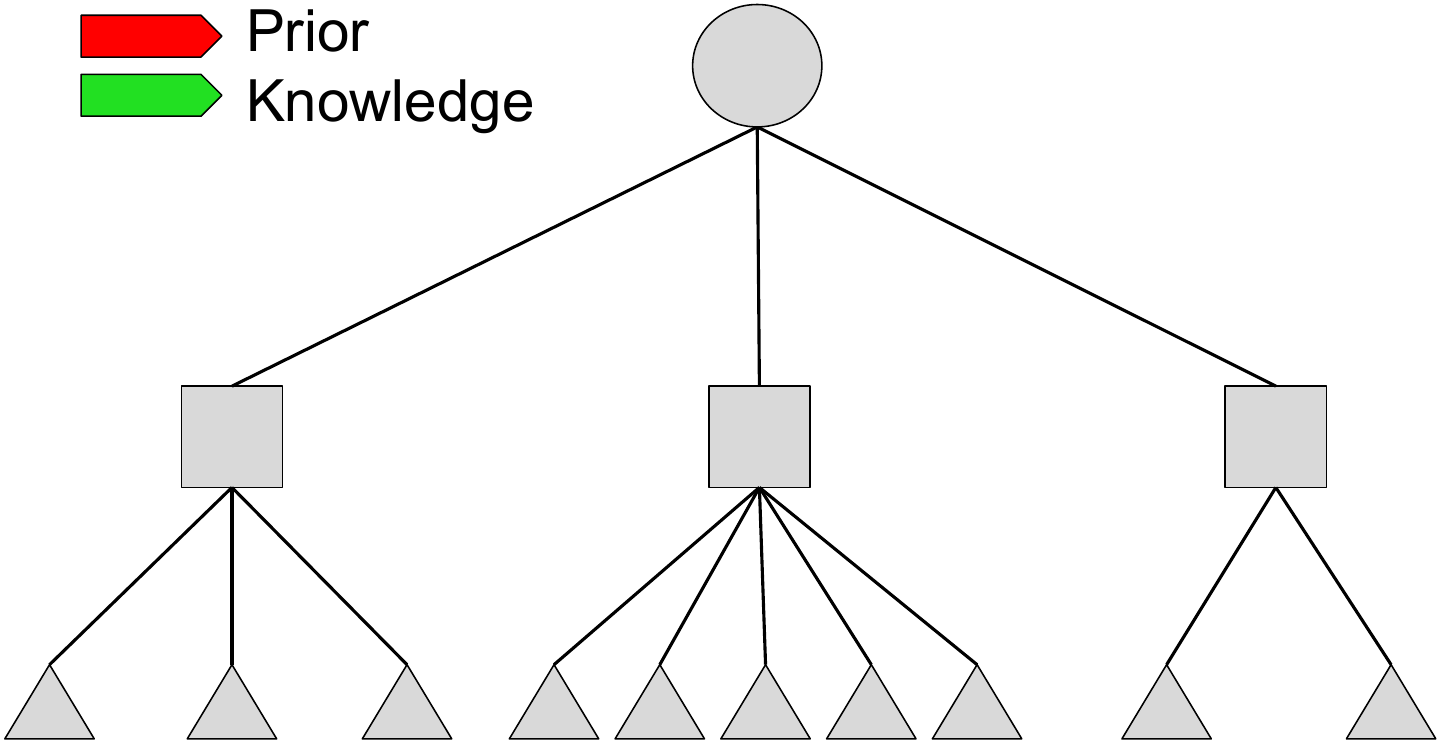}
         \caption{Unannotated project.}
         \label{fig:blank_graph}
     \end{subfigure}
     \hfill
     \begin{subfigure}[b]{0.45\textwidth}
         \centering
         \includegraphics[width=0.95\textwidth]{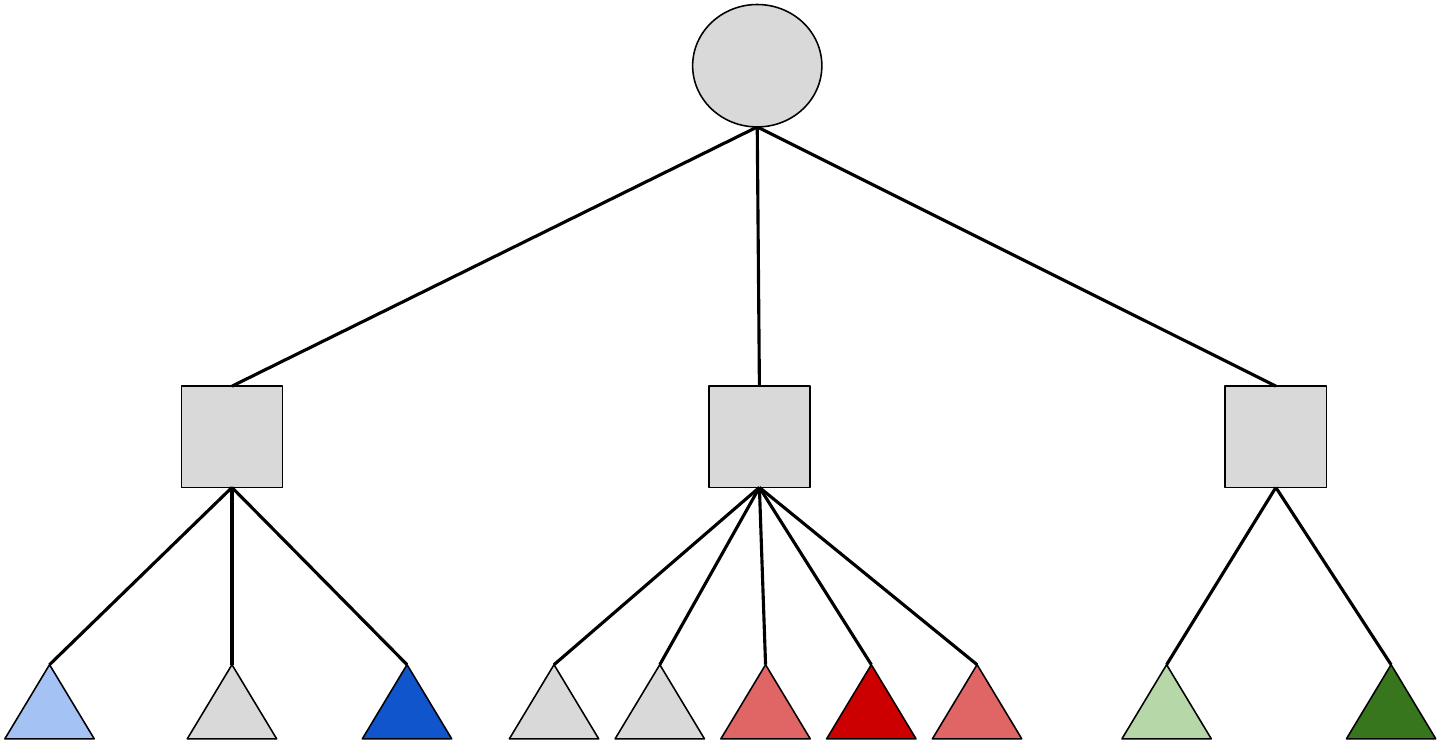}
         \caption{Annotated files.}
         \label{fig:annotated_nodes}
     \end{subfigure}
     \\[1.5ex]
      \begin{subfigure}[b]{0.45\textwidth}
         \centering
         \includegraphics[width=0.95\textwidth]{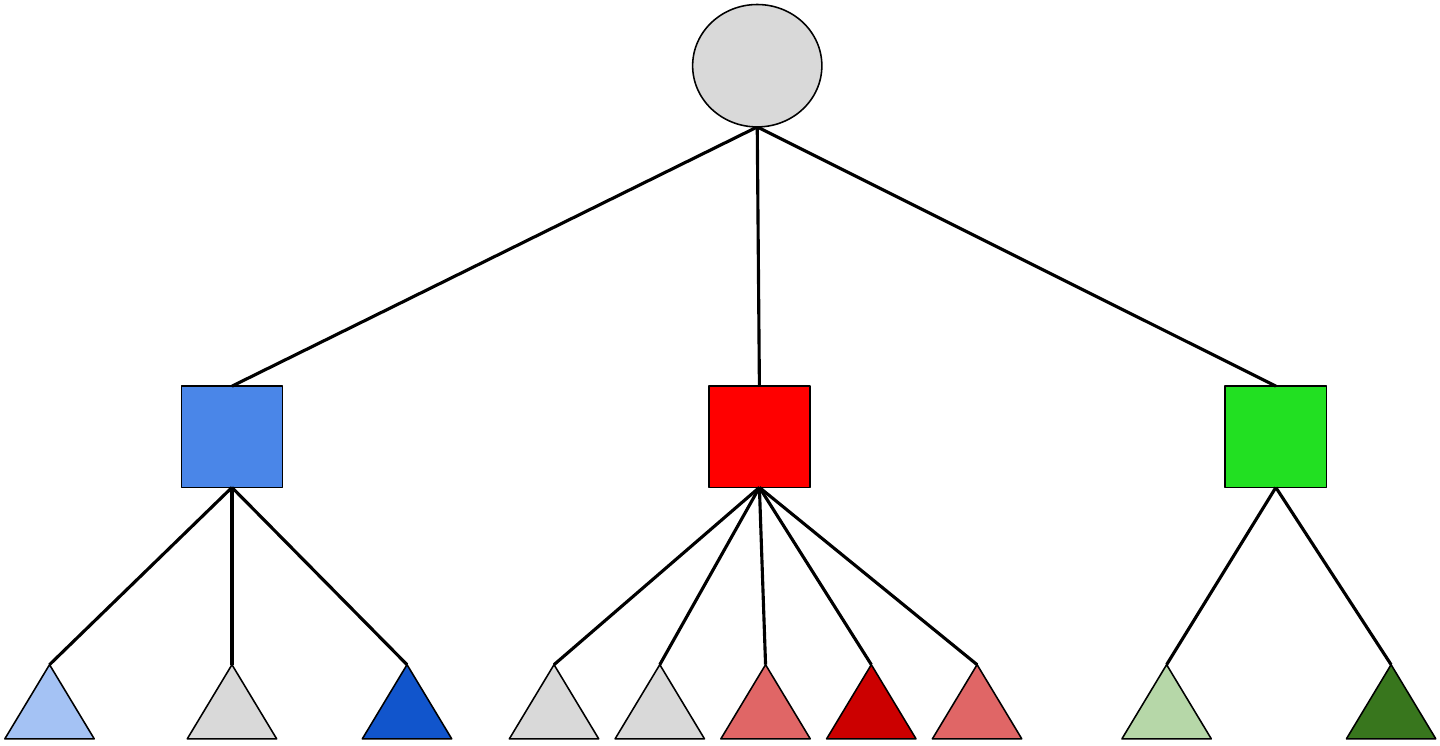}
         \caption{Package-level aggregation.}
         \label{fig:classified_nodes}
     \end{subfigure}
    \hfill
    \begin{subfigure}[b]{0.45\textwidth}
         \centering
         \includegraphics[width=0.95\textwidth]{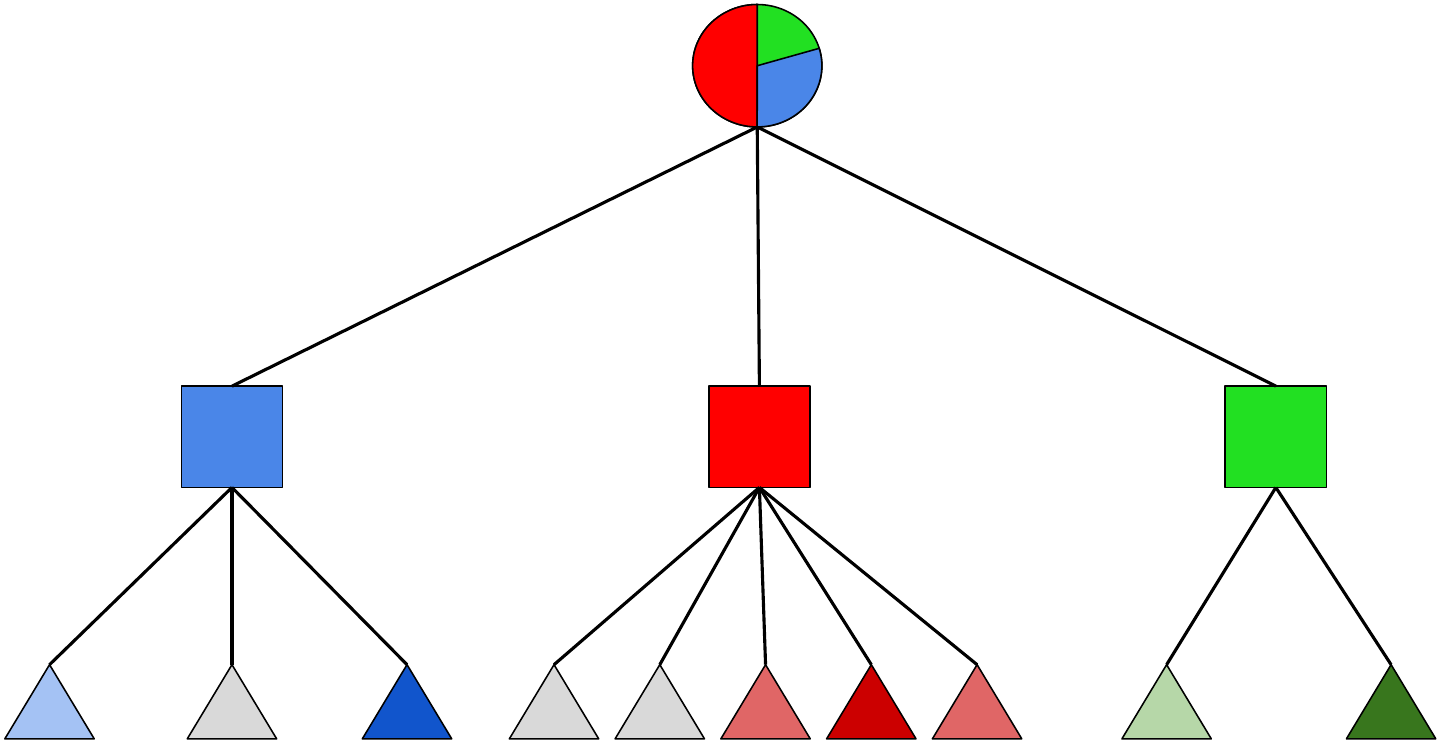}
         \caption{Project-level aggregation.}
         \label{fig:components}
     \end{subfigure}
    \caption{Multi-granular annotation steps. The approach does not use the prior knowledge. Some nodes might not be labelled. Files are triangles, squares are packages, and the circle is the project as a whole. Colour is the label.}
    \label{fig:multi_granular_steps}
\end{figure}

Our paper proposes a weak labelling approach for annotating source code files in a code base. We use this file-level annotation strategy to aggregate annotations at different levels, including package-level and project-level, resulting in the ability to do multi-granular annotations. Figure~\ref{fig:multi_granular_steps} shows a case of how this approach works using an example project: we assume that the project has existing labels (e.g., the `Prior Knowledge' at the top left) that developers assigned to the project. Using a weak labelling approach, it is possible to assign labels to each file (Figure~\ref{fig:annotated_nodes}), and those file-level labels can be lifted to annotate the packages containing the annotated files (Figure~\ref{fig:classified_nodes}). All the annotated packages, in turn, can be used to generate labels for the project as a whole, potentially augmenting the existing, pre-defined labels with new labels extracted from the working code (Figure~\ref{fig:components}).

Figure~\ref{fig:multi_granular_steps} shows that multi-granular annotations can help developers to comprehend the software system's semantic content better and to quickly locate and understand the functionalities of each area in a repository (\eg \lbl{Networking}, or \lbl{Database}).
Moreover, the different levels of granularity can facilitate the identification of new labels not only for the project as a whole but also for the semantic subcomponents that have the potential for reuse.

An automatic method for creating weak labels can be useful and has already produced promising outcomes when used as a source of weak supervision~\cite{zhang2022weak}. As shown in various domains (in computer vision tasks~\cite{khoreva2017weak, papandreou2015weakly}, or natural language tasks~\cite{mekala2020meta, mekala2021coarse}), weak and noisy labels can achieve good results in scarce data, or no annotated data. \revision{It is important to notice that our approach does not perform classification: instead, we focus on the prior step, creating annotated data.}

To evaluate the effectiveness of our approach in performing multi-granular annotations, we used a combination of automated metrics and human annotators. We assessed the ability to correctly annotate at all levels of granularity, the confidence of the annotations, and the number of items that cannot be annotated. 

For the purpose of this work, we are interested in answering the following research questions:

\smallskip

\begin{quote}
    \textit{\textbf{RQ1}: To what extent is weak labelling effective in capturing the semantic content of files for annotation purposes?}
\end{quote}

\begin{quote}
    \textit{\textbf{RQ2}: Does aggregation at the package-level provide an effective means for capturing the content of packages?}
\end{quote}

\begin{quote}
    \textit{\textbf{RQ3}: Does aggregation at the project-level provide an effective means for capturing the content of projects?}
\end{quote}

\begin{quote}
    \textit{\textbf{RQ4}: Can file-level annotations allow the discovery of new topics within projects?}
\end{quote}

This paper is a major extension of our preliminary work~\cite{sas2023weak}, which was focused on three projects only and performed an exploratory analysis of the feasibility of our approach. Besides expanding the dataset, this work evaluates different methods to annotate the data; it also does package-level annotation, and we perform an extended automatic analysis and human evaluation. 

The contributions of this paper are summarized as follows:
\begin{itemize}
    \item A scalable approach based on weak labelling to automatically annotate source code files; 
    \item A framework for multi-granular labelling of software projects, which will allow developers to comprehend the code at different granularities;
    \item A dataset to train models for file-level software classification.
\end{itemize}

\color{black}
The paper is structured as follows: in Section~\ref{sec:motivating}, we present a motivating example for this work. Section~\ref{ref:background}, offers some background knowledge for techniques and methods used in this work. Section~\ref{sec:proposed} presents the proposed approach, which is evaluated using the methods explained in Section~\ref{sec:eval}. The evaluation results are shown in \ref{sec:results}. The discussion of the results is presented in Section~\ref{sec:discussion}, and an overview of possible uses of our work is discussed in Section~\ref{sec:uses}. We present the threats to validity in Section~\ref{sec:threats}, and in Section~\ref{sec:sota}, we discuss previous work related to our own. Lastly, Section~\ref{sec:conclusions} presents the conclusions and future works.

\section{Motivating Example}
\label{sec:motivating}

A program is not a single, monolithic piece of code that performs a single function: instead, it is a set of modules, each contributing differently and interacting with others to create different functionalities in the software.
Current solutions ignore this aspect and classify the software as a whole, using proxies like the \README files. For example, if we consider a project like Weka~\footnote{\href{https://github.com/Waikato/weka-3.8}{https://github.com/Waikato/weka-3.8}}, an ML desktop application and library, or Pumpernickel~\footnote{\href{https://github.com/mickleness/pumpernickel}{https://github.com/mickleness/pumpernickel}}, a small UI library, we can see the downsides of this approach. The \README files give pieces of information that are unrelated to the project content and its application.

\begin{figure}[htb!]
    \centering
        \begin{subfigure}[b]{\textwidth}
         \centering
         \includegraphics[width=0.49\textwidth]{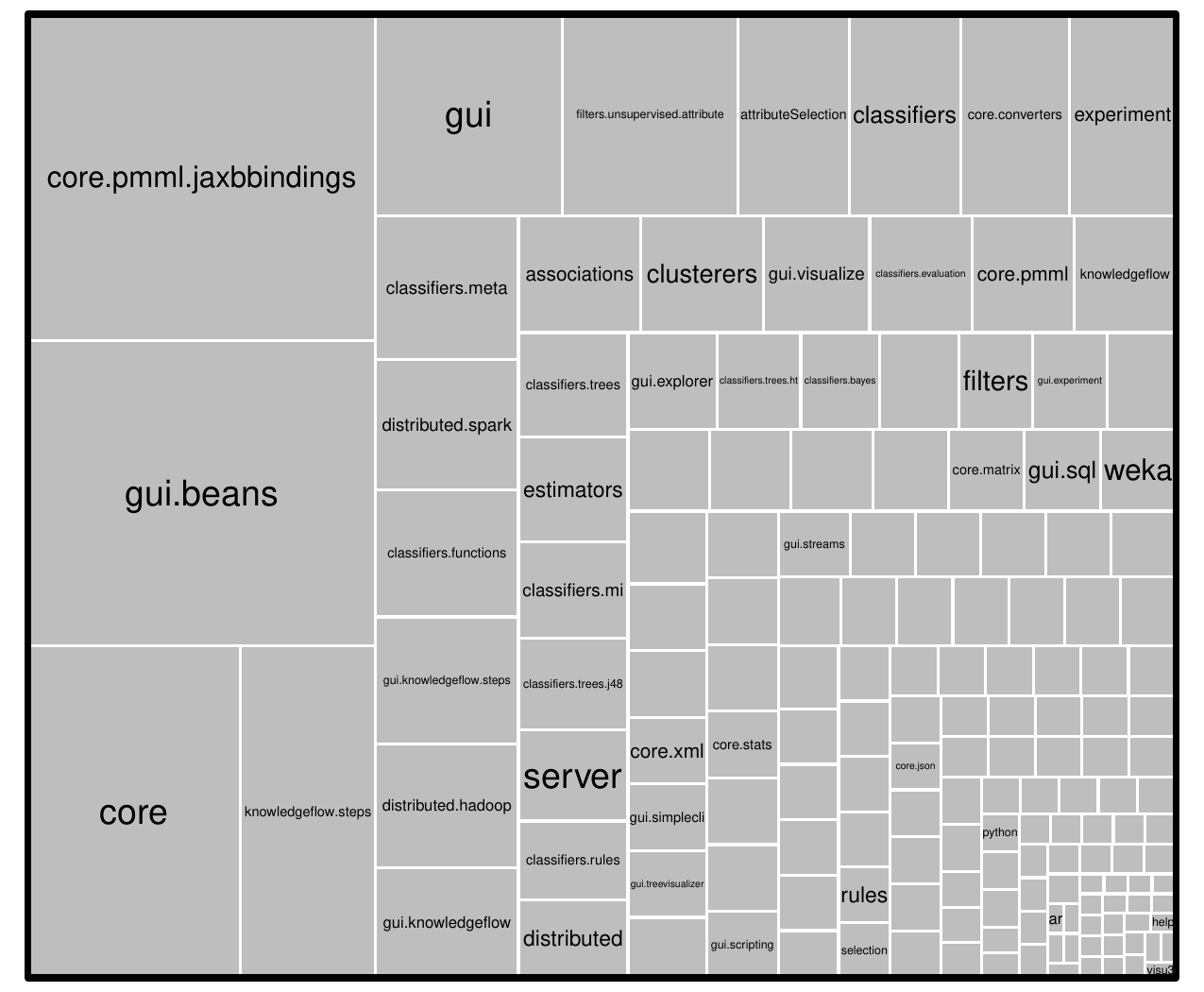}
         \centering
         \includegraphics[width=0.49\textwidth]{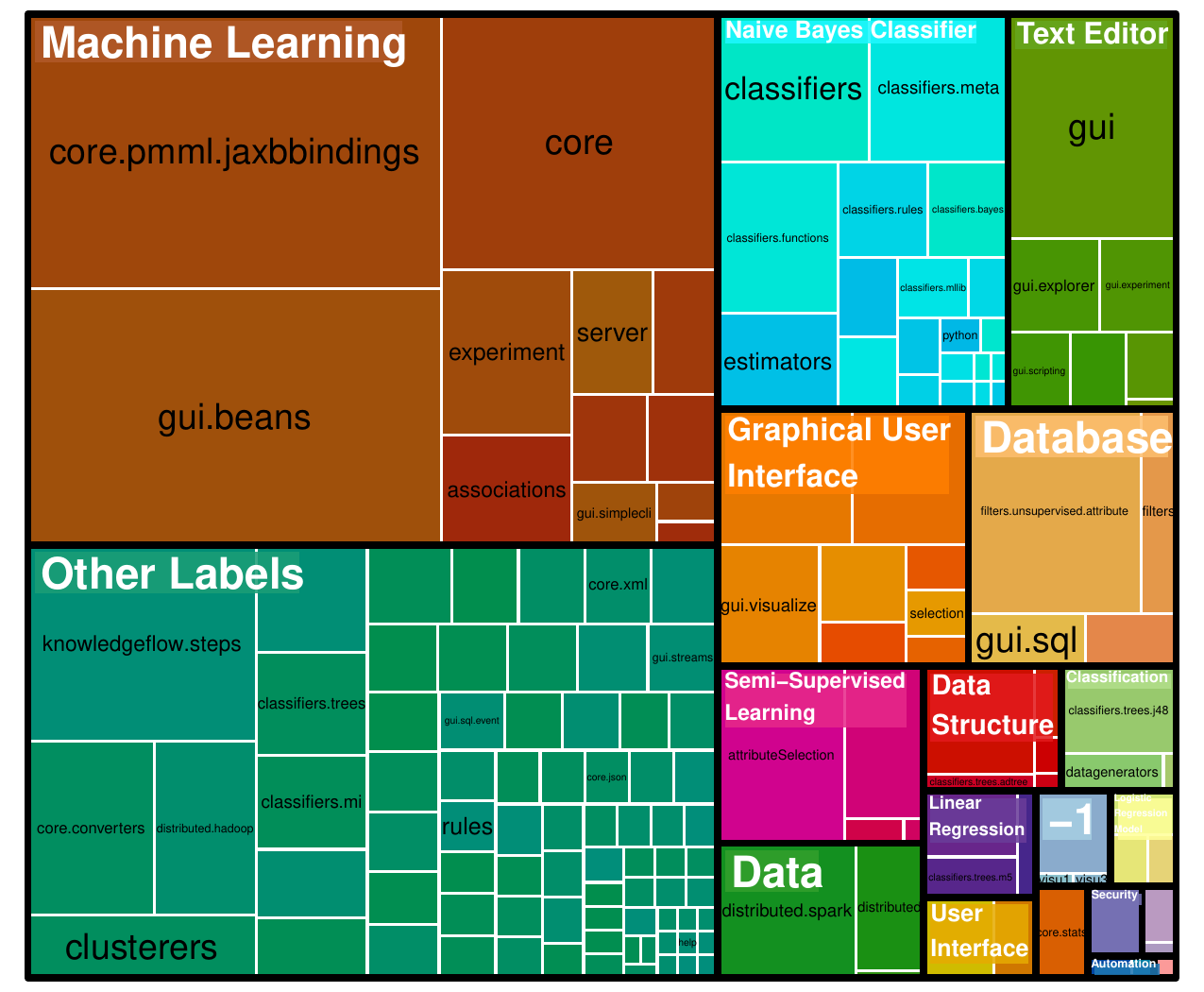}
         \caption{Weka packages, before and after annotation.}
         \label{subfig:weka_treemap_20}
     \end{subfigure}
         \centering
    \begin{subfigure}[b]{\textwidth}
         \centering
         \includegraphics[width=0.49\textwidth]{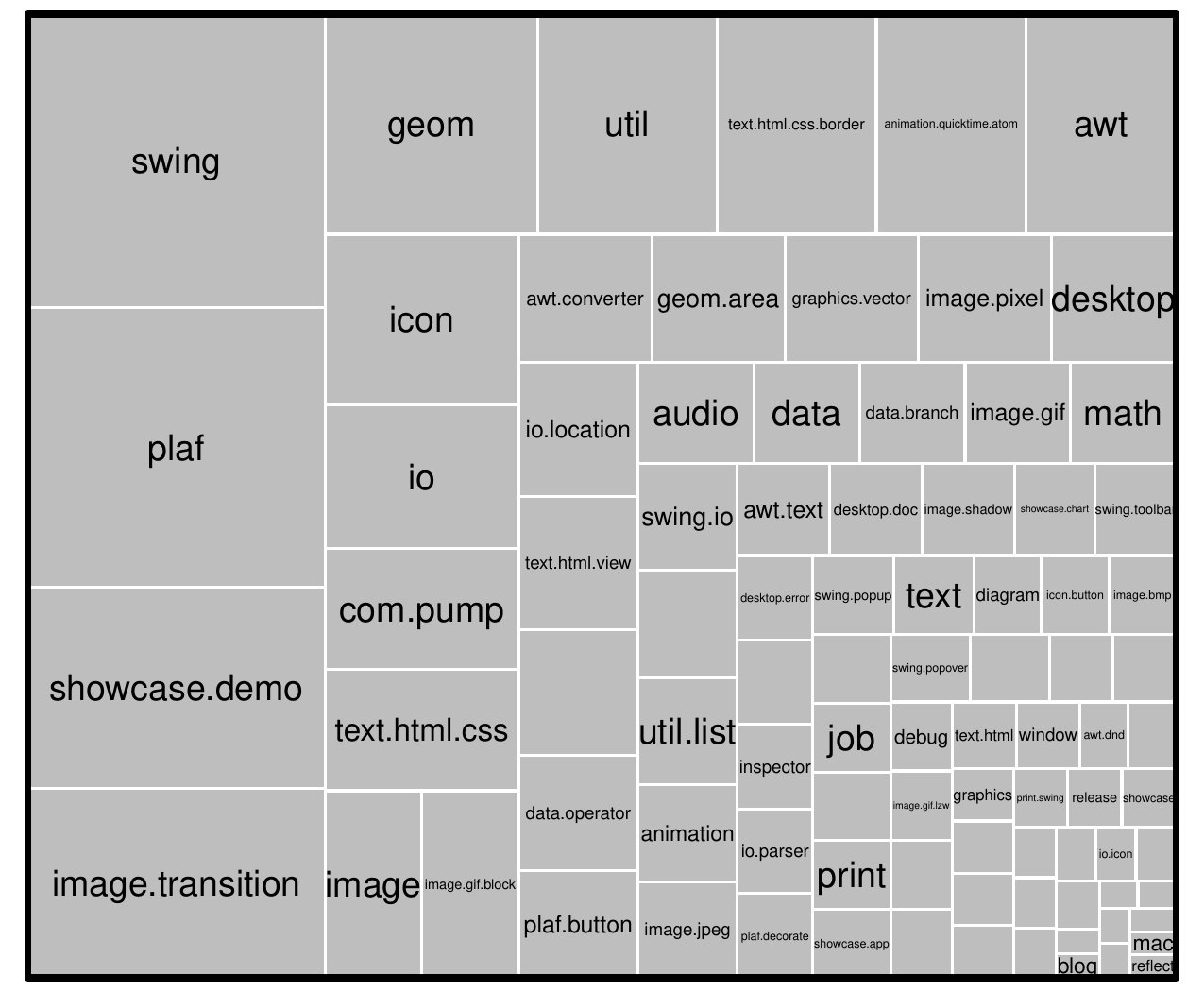}
         \centering
         \includegraphics[width=0.49\textwidth]{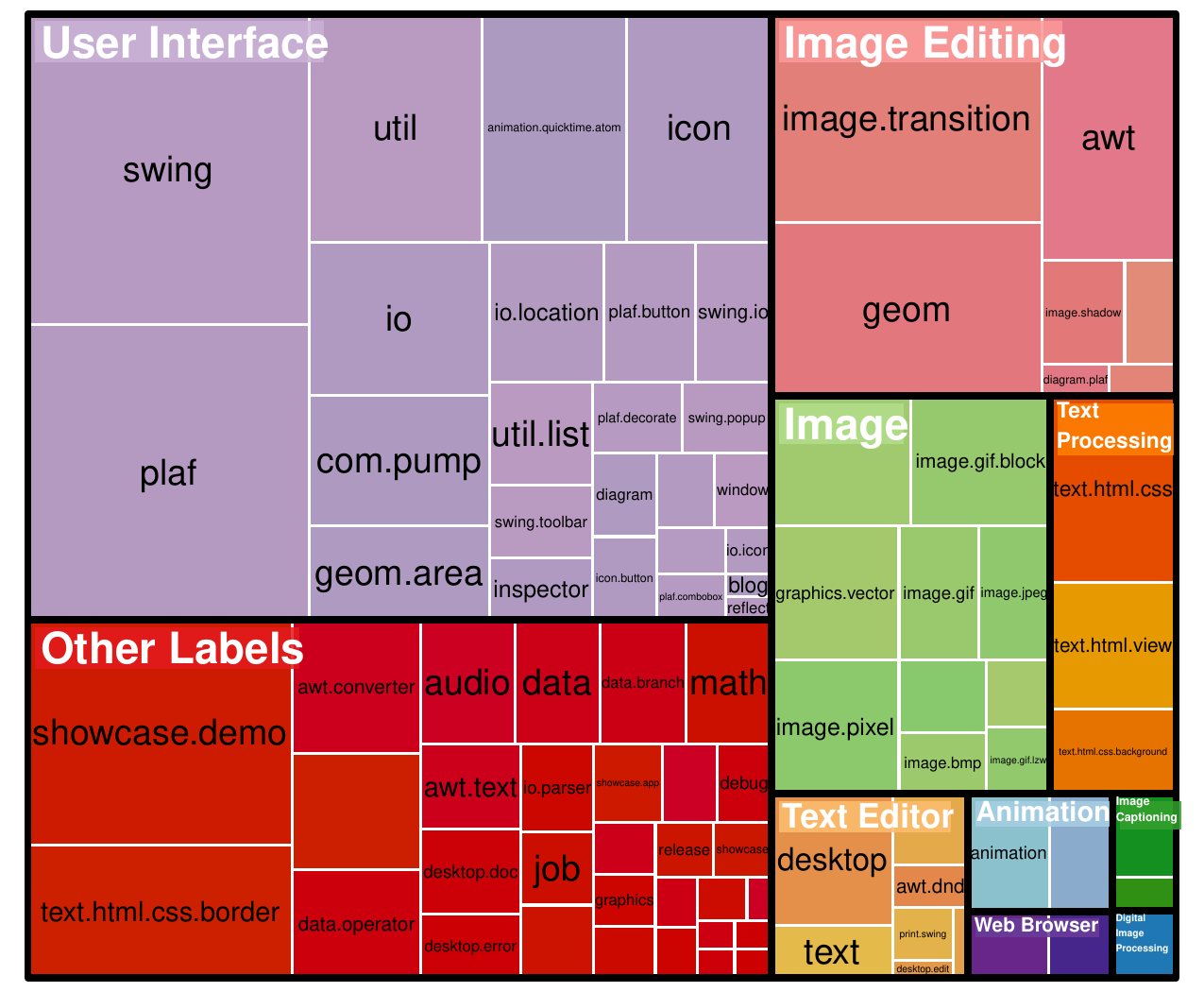}
         \caption{Pumpernickel packages, before and after annotation.}
         \label{subfig:pumpernickel_treemap_10}
     \end{subfigure}
    \caption{Package annotations for Pumpernickel and Weka. Labels are identified using the approach proposed in this work. Texts in black are the package names, and texts in white are the labels. Colour shades are for the same labels. The \lbl{Other Labels} encompasses labels not at the top.}
    \label{fig:package_labels}
\end{figure}

If we look at the Topics in GitHub, Weka reports only \lbl{Machine Learning} as a label; similarly, the Pumpernickel project only lists the \lbl{UI} topic. However, using the project's content, one can find more information for inferring labels. Using our approach, along with the \lbl{Machine Learning} label assigned by the developers, we can identify more specific instances of ML (\eg \lbl{Na\"{i}ve Bayes Classifier}) and parts that one might not be immediately aware of, like the \lbl{Graphical User Interface} parts. The composition of labels for the Weka project can be viewed in Figure~\ref{subfig:weka_treemap_20}, where packages are annotated with a label. Packages have been annotated as \lbl{Other Labels} when none of the most likely labels (in Figure~\ref{subfig:weka_treemap_20} we display 20) can be applied. Lastly, some wrong classifications are visible, like the \texttt{weka.gui} package being labelled as \lbl{Text Editor}. 

The same file-level annotations can be used also to annotate the Pumpernickel project (Figure~\ref{subfig:pumpernickel_treemap_10}), and to complement the labels provided by the developers. As visible from the figure, we can identify parts responsible for \lbl{Image Editing} and \lbl{Text Editing}. These labels allow developers to gain an overview of the content of the projects quickly and reduce the time required to familiarize themselves with new unknown projects. Furthermore, identifying modules responsible for specific tasks can be helpful for software reuse.

In the remainder of this paper, we report the methodology used to extract file-level labels and annotate packages and projects, together with our results.

\section{Background}
\label{ref:background}

In this section, we provide a detailed description of the techniques that form the basis of our approach, particularly `weak labelling' and `keyword extraction'. Weak labelling enables us to annotate the data with minimal effort, while keyword extraction allows us to extract domain-specific knowledge used in the annotation process. 

\subsection{Weak Labelling} 
Machine learning techniques have revolutionized research in many areas; however, these models depend on access to high-quality labelled training data. Yet, collecting human annotations is not always feasible and might be impractical (\eg annotating pixels in images for pixel-level semantic segmentation~\cite{papandreou2015weakly}, or natural language classification~\cite{mekala2020meta}).

Weak supervision~\cite{zhang2022weak} is a growing area of research in machine learning that aims to train machine learning models using incomplete, noisy, or imprecise methods created by weak labelling. Weak labelling uses heuristics, rules, or domain-specific knowledge to automatically assign labels to the observed data based on their characteristics rather than relying on manual annotation. 

The pipeline for generating weak labels consists of using different Labelling Functions (LFs), each with a different source of supervision. However, there is also a need to combine these LFs as they might have different characteristics.

One approach combines these LFs using a label model~\cite{ratner2019labelmodel}, a weighted ensemble of the LFs, where the weights are learned in a probabilistic fashion using graphical models. One limitation is that the label models are currently only suited for a single prediction per LF; as a solution, we will use simple approaches like voting.

\begin{figure}
    \centering
    \begin{subfigure}[b]{\textwidth}
         \centering
         \includegraphics[width=0.95\textwidth]{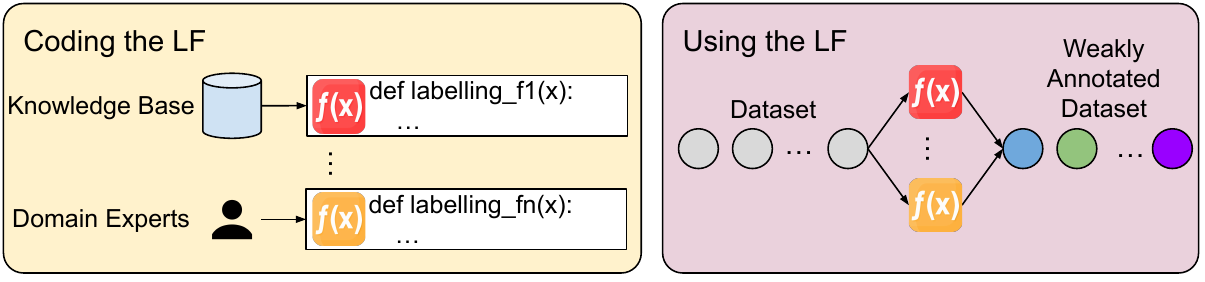}
         \caption{Example of a weak labelling pipeline.}
         \label{fig:labelling_example}
     \end{subfigure}
     \par\bigskip 
     \begin{subfigure}[b]{\textwidth}
         \centering
         \includegraphics[width=0.95\textwidth]{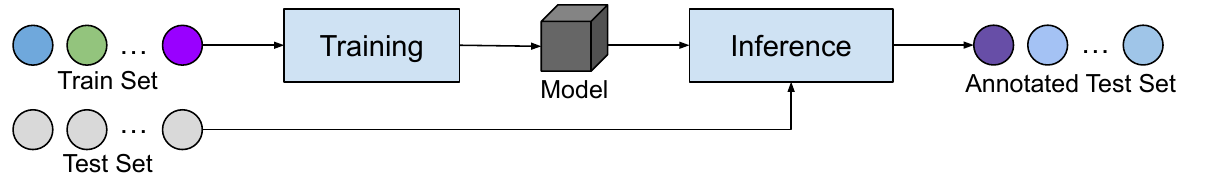}
         \caption{Example of a classification pipeline.}
         \label{fig:classification_example}
     \end{subfigure}
    \caption{Differences between labelling and classification.}
    \label{fig:labelling_vs_classification}
\end{figure}

\revision{While the annotation task might seem like classification, in weak labelling, there is no training and no inference. Therefore, instead of using annotated examples and learn the parameters of a function defining a separating hyperplane, external knowledge is combined with heuristics (in the form of LF) to assign a noisy label to an example (Figure~\ref{fig:labelling_vs_classification}).}

\subsection{Keyword Extraction}

Keyword extraction is a critical step in text mining, particularly when the number of available documents or domains grows. Given the sheer volume of documents, it is impractical for a user to read them all in detail. The objective of keyword extraction is to identify the words that most effectively represent the document~\cite{firoozeh2020keyword}.

There are several approaches to extracting keywords, including simple statistical methods, linguistic models, and machine learning models~\cite{bharti2017keywordsurvey}. However, since machine learning models require annotated data and linguistic models rely on external knowledge, we focus on statistical methods in this study, specifically for domain-specific documents.

We use a state-of-the-art unsupervised statistical approach called \yake~\cite{campos2020yake} to extract keywords in this work. This algorithm tokenizes the text and removes English stopwords. It then calculates various statistics for each term, such as frequency, co-occurrences, position in the text, and the number of sentences it appears in. To identify $n$-gram keywords, a sliding window approach is employed. The final score of each keyword is a product of the scores of each term belonging to the keyword normalized by the keyword frequency. This method effectively enables us to extract relevant and informative keywords from domain-specific documents without prior knowledge.

\subsection{Word Embeddings}
Word embeddings are a popular technique in natural language processing (NLP) and machine learning to represent words as numerical vectors in a high-dimensional space. Word embeddings are typically learned from large amounts of text data using neural network models that model the probabilities of words in textual data. As such, they capture the intricate semantic and syntactic relationships between words.
Word embeddings are helpful as they can be trained on some data and then applied for other downstream tasks. In our case, we use word embeddings to model textual data to measure similarities between two texts.

One example of such models is Word2Vec~\cite{Mikolov2013w2v}, \revision{a neural network model} that learns the embeddings of words by using the context (e.g., their neighbouring words) in which the word occurs. These embeddings can then be averaged to compute the embedding of a sentence. One adaptation of this approach specifically for the software engineering domain is the Stack Overflow embeddings SO-W2V~\cite{efstathiou2018sow2v}.

One issue with current Word2Vec approaches is their limited vocabulary, making it impossible to model words that have not been seen during training; one way to address this issue is to use subword information like $n$-grams. On the other hand, FastText~\cite{bojanowski-etal-2017-enriching} embeddings use $n$-grams as their building blocks and average their embeddings to create the word and sentence embeddings. 

While models that use subword information effectively solve the out-of-vocabulary issue, they do not consider the specific context in which a word appears. Contextualized Language Models (LMs) like BERT~\cite{devlin-etal-2019-bert} create word vectors that also contain the information of the context, making it easier to disambiguate the meaning of a word.

Lastly, there are also code-specific LM~\cite{uri2019code2vec, feng2020CodeBERT, allal2023santacoder}; however, in this paper, we are interested in the natural language meaning of the terms present in the code, rather than their code-specific syntactic and semantic information. On the other hand, code-specific LMs are generally trained for code completion and generation tasks: as a result, even using them to extract features will require some fine-tuning. Nevertheless, adaptations of these models could be used in future work for training models to perform classification using the annotation created from this work. 

\section{Methodology}
\label{sec:proposed}

\begin{figure}[htbp!]
    \centering
    \includegraphics[width=\textwidth]{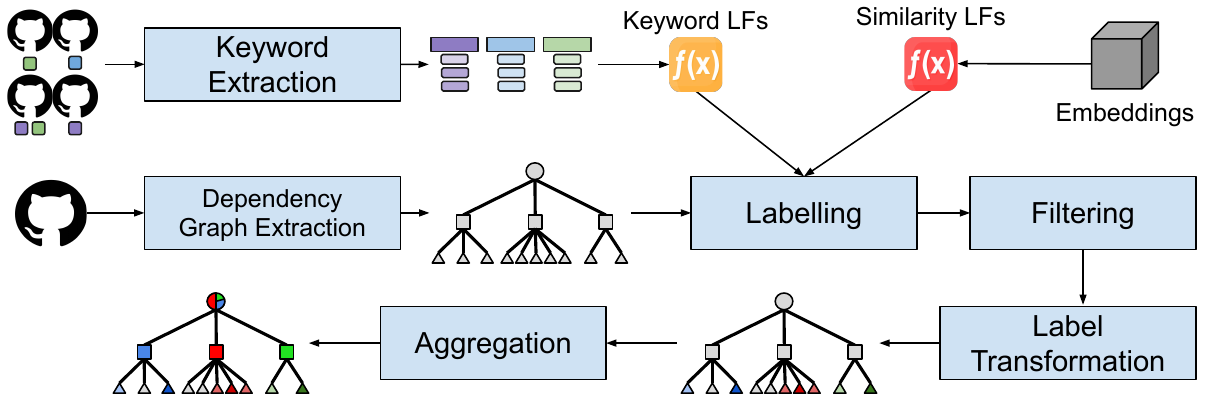}
    \caption{Pipeline for the proposed approach.}
    \label{fig:pipeline}
\end{figure}

Our methodology is illustrated in Figure~\ref{fig:pipeline}, which shows the various steps of the pipeline. In the following sections, we provide a detailed description of each step. Furthermore, to promote reproducibility and enable future research, we have made our code\footnote{\href{https://github.com/SasCezar/CodeGraphClassification}{https://github.com/SasCezar/CodeGraphClassification}} and data\footnote{\href{https://zenodo.org/record/7943882}{https://zenodo.org/record/7943882}} publicly available.

\subsection{Dataset}
The dataset adopted for our experiments is a subset of our previous work GitRanking~\cite{sas2022gitranking}; however, we restrict our analysis to solely Java projects. 
The dataset uses a subset of GitHub Topics that have been manually checked to ensure their relevance as software application domains, including categories such as \lbl{Networking} and \lbl{Database} while excluding others like \lbl{Google}. Furthermore, the dataset's labels are linked to Wikidata~\cite{vrandecic2012wikidata}, an external knowledge base for disambiguation purposes. 
The subset used for this work contains only the Java projects of the dataset (an overall 2,795 projects) accompanied by a set of 267 unique labels. A subset of such labels can be seen in Table~\ref{tab:topic_rank_list}: as seen from the excerpt, all the labels are linked to a software-related application domain~\cite{glass1995contemporary}.

\begin{table}[htb!]
\centering
\caption{Subset of labels used to annotate the projects in our dataset.}
\begin{tabular}{c}
\toprule
\textbf{Label}  \\
\midrule
Machine Learning  \\
Graphical User Interface   \\
Database  \\
Animation    \\
Linear Regression  \\
Software Engineering \\
\textbf{\dots} \\
\bottomrule
\end{tabular}
\label{tab:topic_rank_list}
\end{table}

\subsection{Dependency Graph Extraction}
For the multi-granular approach to work, we need to know the structure of each project. We used the Arcan tool~\cite{fontana2017arcan} to extract the dependency graph, obtaining the complete set of nodes and edges describing the dependencies between classes and packages. Furthermore, Arcan also extracts dependencies between files, which can be used as extra information in future works.

\subsection{Keyword Extraction}
\revision{The idea behind using keyword extraction algorithms is to identify the most important terms in each project, and then assign these terms to the developers' assigned labels.} In this work, we used a state-of-the-art unsupervised statistical approach called \yake~\cite{campos2020yake} to extract the keywords. 

In our case, a document is a software project, and we extracted the project's content using the file names. With Java syntax, we used a simple camel case tokenizer to split the composed words into individual terms. This approach, compared to extracting keywords from the source code, reaches a high throughput as less text is being analysed, while maintaining enough information to describe the content~\cite{nemitari2016semantic} and reduced noise from non-informative identifiers in the code (\eg like \textit{i}, abbreviations, or typos). \revision{In Table~\ref{tab:project_keywords} we can see some keywords for the Weka and Pumpernickel projects.}

\begin{table}[!htbp]
\centering
\caption{Project keywords extracted by \yake.}
\begin{tabular}{l|l}
\toprule 
\textbf{Weka} & \textbf{Pumpernickel} \\
\midrule
classifier &  gif \\
data &  css \\
jaxbbindings & image \\ 
tree &  button \\
regression &  shape\\
bayes & renderer \\
\textbf{\dots}  & \textbf{\dots} \\
            \bottomrule
\end{tabular}
\label{tab:project_keywords}
\end{table}


\revision{The project extracted keywords are then assigned to all the labels that the project has been annotated by the developer on GitHub.} 

Once we extracted the keywords for all labels, we compute a weight for each keyword; our choice to assign weights to keywords was based on TF-IDF. The \textit{document} is the label, the extracted \textit{terms} are the words, and the \textit{frequency} of the keyword is the number of occurrences in the documents annotated with that label. 

We adopted this weighing as a mechanism to reduce terms that appear in multiple labels, as each keyword for a project is assigned to all the labels the project is annotated with; therefore, some keywords are likely to appear in more labels but will have a higher weight in the correct label. In Table~\ref{tab:label_keywords}, we see an example of keywords for two labels. The table also shows some unrelated keywords (\eg \lbl{Database} in the  \lbl{Machine Learning} label) that are present in the lists: as mentioned above, this is because it is not possible to separate what label each keyword belongs to when multiple labels are available per project.

\begin{table}[!htbp]
\centering
\caption{Subset of keywords of two labels with the respective TFIDF.}
\begin{tabular}{lc|lc}
\toprule 
\multicolumn{2}{c|}{\textbf{Machine Learning}} & \multicolumn{2}{c}{\textbf{Graphical User Interface}} \\
\midrule
\textit{Keyword}            & \textit{TFIDF}           & \textit{Keyword}     & \textit{TFIDF}     \\
\midrule 
ELKI  & 0.67                  &             Editor  & 0.41           \\
Clustering & 0.06            &             Swing & 0.17         \\
Classification & 0.03       &  Refactoring & 0.15           \\        NLP     &  0.02                  &       Scene & 0.12          \\             Database &  0.02                 &      Widget     &  0.08          \\
\textbf{\dots} & \textbf{\dots} & \textbf{\dots} & \textbf{\dots} \\
            \bottomrule
\end{tabular}
\label{tab:label_keywords}
\end{table}

\subsection{Labelling}

Our \revision{weak labelling} approach employs two distinct types of LFs for annotation. The first type of LF is based on keyword matching (\ref{sec:label-KW}), while the second type uses semantic features (\ref{sec:label-sim}). 
Our LFs do not return a single prediction; instead, they produce, as an output, a vector that represents the probability distribution over a set of $m$ variables (in our case, the labels). We discuss the two types of LFs below.

\subsubsection{Keyword-based Labelling Functions}
\label{sec:label-KW}

The keyword labelling function uses the keywords extracted from the file names and checks if the analyzed documents contain these keywords. 

For example, for the label \lbl{Machine Learning}, we can see some terms in Table~\ref{tab:label_keywords}. 
For Weka, the file \texttt{../classifiers/meta/ClassificationViaClustering.java}, if we use the name, will result in the document with the terms: \textit{classifiers, meta, classification, via, clustering}. Combining the terms in the document, and the ones in the \lbl{Machine learning} label, an overall probability of $0.0825$ will result, being the third most likely label for the file \revision{(see Table~\ref{tab:keyword_label})}.

\begin{table}[htbp!]
    \centering
    \caption{Probabilities for the top labels using the \textit{keyword}-based LF on the name. The file is \texttt{../classifiers/meta/ClassificationViaClustering.java} and belongs to the Weka project.}
\begin{tabular}{lc}
\toprule
                 \textbf{Label} &     \textbf{Prob} \\
\midrule
        Naive Bayes Classifier &       0.1192 \\
Classification &       0.1100 \\
      Machine Learning &       0.0825 \\
           Data Mining &       0.0550 \\
      Data Analysis &       0.0550 \\
  \dots & \dots \\
\bottomrule
\end{tabular}
    \label{tab:keyword_label}
\end{table}

Formally, the label scores of a node (file) $n$ given a label $l$ are defined as:  
\begin{equation}
    LS_{n,l} = \sum_{t \in terms(n)}  freq(t, n) \times weight(t, l) 
\end{equation}
where:

\begin{itemize} [noitemsep]
    \item $terms(n)$: gives us all the terms in the source file $n$;
    \item $freq(t, n)$: represents the frequency of the term $t$ in the file;
    \item $weight(t, l)$: is a weight computed for each keyword (in our case, TF-IDF).
\end{itemize}

Lastly, we normalize the label score by dividing each score by the sum of the scores of all labels. 

We apply the \textit{keyword}-based LFs on two different modalities: the file \textbf{name} itself and the \textbf{identifiers} in the source file. A Java identifier can be a class, method or variable name. \revision{We use the tree-sitter\footnote{\href{https://github.com/tree-sitter/tree-sitter}{https://github.com/tree-sitter/tree-sitter}} library for the parsing of source code files and extract the identifiers.}

\subsubsection{Similarity-based Labelling Functions}
\label{sec:label-sim}
For the \textit{similarity}-based LFs, the labels' distribution is computed using the semantic similarity between the label and source code file name. An example can be seen in Table~\ref{tab:similarity_sim}, with the probability (normalized similarity) for various labels for the file \texttt{../classifiers/meta/ClassificationViaClustering.java} in the Weka project.

The label score of a node with name $n$ and a label $l$ is defined as:  
\begin{equation}
    LS_{n,l} = sim(n, l) 
\end{equation}
for a given semantic similarity function $sim$. Our choice for $sim()$ is the cosine similarity. Since the cosine similarity is bounded between $[-1, 1]$, we normalize the vector by summing the absolute value of the minimum score and then performing normalization, turning it into a probability vector, with the values in the $[0, 1]$ range, and $norm=1$.

 We use \textbf{fastText}, \textbf{BERT}, and \textbf{W2V-SO} embeddings models on the name.
 
\begin{table}[htbp!]
    \centering
    \caption{Probabilities for the top labels using the W2V-SO model. The file is \texttt{../classifiers/meta/ClassificationViaClustering.java} and belongs to the Weka project.}
\begin{tabular}{lc}
\toprule
                 \textbf{Label} &     \textbf{Prob} \\
\midrule
        Classification &       0.5327 \\
Naive Bayes Classifier &       0.4766 \\
      Cluster Analysis &       0.4205 \\
           Data Mining &       0.3364 \\
      Machine Learning &       0.2243 \\
  \dots & \dots \\
\bottomrule
\end{tabular}
    \label{tab:similarity_sim}
\end{table}

\subsection{Filtering}
The LFs we used can always annotate a file, even when highly uncertain, making the annotation very noisy. The noise is expressed as a very uniform distribution in the probability vector. We adopt the Jensen–Shannon Distance (JSD)~\cite{endres2003metric} to measure how close the prediction is to the uniform distribution. The JSD is a symmetric and bounded metric to compute the distance between two probability distributions. The JSD is the square root of the average of the forward and backward Kullback–Leibler divergence~\cite{kullback1951information}, a distance measure between distributions. The JSD ranges from 0 (the distributions are identical) to 1 (dissimilar).
We test different thresholds to mark the files with a JSD lower than a threshold as `\textit{unannotated}'. Along with no filtering, two thresholds (\ie $0.25$ and $0.5$) were tested in this work. 

\begin{figure}[htbp!]
    \centering
    \includegraphics[width=.8\textwidth]{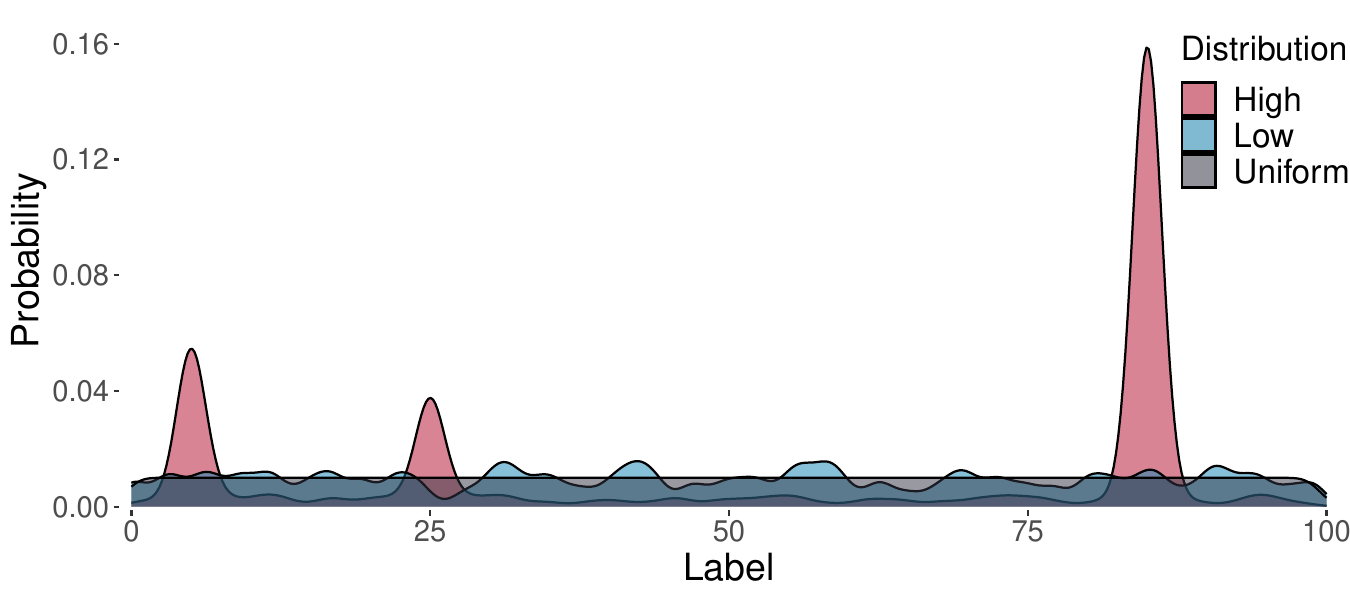}
    \caption{Example of two distributions that have different JSD w.r.t. the uniform distribution (grey). }
    \label{fig:example_jsd}
\end{figure}

Figure~\ref{fig:example_jsd} presents a visual example of the JSD and why it is an effective filtering approach. When measured against the uniform distribution (grey), the high JSD distribution (red) exhibits a high probability for a few labels, whereas the low JSD one (blue) has low probabilities overall. The thresholds effectively help to select the probability peaks with more or less confidence.

\revision{In the case of the annotations examples presented in Table~\ref{tab:keyword_label}, and Table~\ref{tab:similarity_sim}, the JSD score is respectively 0.74, and 0.20, therefore the \textit{keyword}-based approach (Table~\ref{tab:keyword_label}) will not be filtered in any case, while the W2V-SO LF annotation will be marked as \textit{unannotated} for both filtering settings.}

\subsection{Label Transformation}

\begin{figure*}[htbp!]
    \centering
    \begin{subfigure}[b]{0.3\textwidth}
         \centering
         \includegraphics[width=0.95\textwidth]{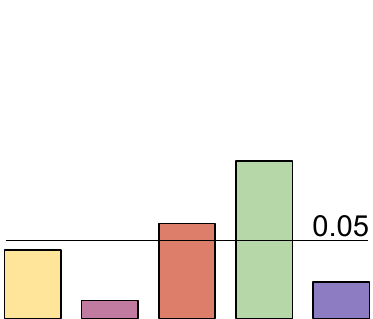}
         \caption{RAW}
         \label{fig:transformation_none}
     \end{subfigure}
     \hfill
     \begin{subfigure}[b]{0.3\textwidth}
         \centering
         \includegraphics[width=0.95\textwidth]{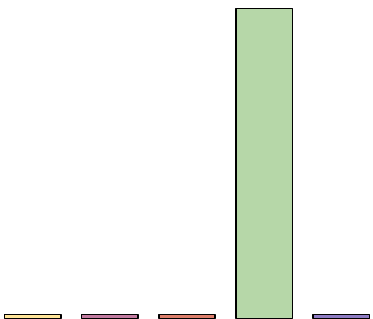}
         \caption{T1}
         \label{fig:transformation_t1}
     \end{subfigure}
          \hfill
      \begin{subfigure}[b]{0.3\textwidth}
         \centering
         \includegraphics[width=0.95\textwidth]{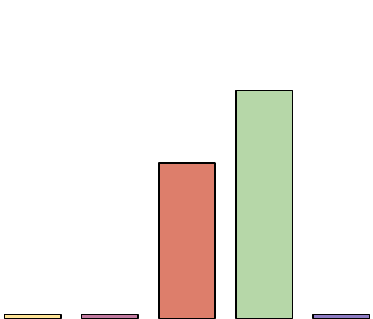}
         \caption{Tp}
         \label{fig:transformation_tp}
     \end{subfigure}
    \caption{Results of applying different transformation functions. Figure~\ref{fig:transformation_none}, is the raw annotation with a line indicating the threshold used for Tp.}
    \label{fig:transformations}
\end{figure*}

The labelling functions used are returning distributions, which are soft labels, meaning each label has a non-zero probability attached: Figure~\ref{fig:example_jsd} shows how the probability varies for each label. Along the raw output without any transformation (\textbf{RAW}), we also investigate different transformations to the distributions to improve the performance (Figure~\ref{fig:transformations}). One obvious transformation is to pick only the highest probability label (\textbf{T1}) as the only label, as displayed in Figure~\ref{fig:transformation_t1}. Another approach is to pick only the labels with a probability higher than a threshold (\textbf{Tp}); we pick $0.05$ as a 12x over the uniform probability to keep only the confident predictions, shown in Figure~\ref{fig:transformation_tp}. The results are normalized to maintain the annotations as probabilities.

\revision{An example of transformation can be seen in Table~\ref{tab:tp_transformation}, where we apply the \textbf{Tp} transformation to the example in Table~\ref{tab:similarity_sim}. As we notice the original \textbf{RAW} probabilities are changed, with an increase for the top labels, and the labels with lower probabilities (from the sixth) are suppressed to 0.}

\begin{table}[htbp!]
    \centering
    \caption{\textbf{Tp} transformation applied to the Weka project file \texttt{../classifiers/meta/ClassificationViaClustering.java}.}
\begin{tabular}{lc}
\toprule
                 \textbf{Label} &     \textbf{Prob} \\
\midrule
        Naive Bayes Classifier &       0.6022 \\
Classification &       0.5588 \\
      Machine Learning &       0.4169 \\
           Data Mining &       0.2779 \\
      Data Analysis &       0.2779 \\
  \dots & 0.0 \\
\bottomrule
\end{tabular}
    \label{tab:tp_transformation}
\end{table}

\subsection{Aggregation}
We obtained the multi-granular annotations (for a package, or the project as a whole) by aggregating the file-level annotations. We chose a na\"{i}ve solution as an initial approach, \revision{the final probability vector for the package, and project are computed using the average over the files vectors}. Figure~\ref{fig:multi_granular_steps} shows an example of aggregation; the information from the files is averaged over for the package-level annotations (Figure~\ref{fig:classified_nodes}) and project-level (Figure~\ref{fig:components}). 

\revision{For the aggregation at the project-level}, we \revision{computed} the mean over all \revision{vectors} of the annotated \revision{files} and picked the top $K$ labels. The resulting labels can be used to evaluate the file-level annotations indirectly, as project-level annotations are the only source of supervision available. 

The package-level annotations were predicted similarly to the project-level; however, we only considered the files that belong to that package when aggregating. \revision{Using the dependency graph we get all the annotated files and average their probabilities vector to get the package vector. Furthermore}, we filtered out the labels not in the top $K$ for the project to reduce noise and avoid having too many labels for a project. Therefore, the main label used for visualization is the label with the highest probability, also in the top $K$ at the project level.
If there is no label in the top $K$ labels in the package, the package is marked as \lbl{Unannotated}.

\subsection{Ensemble}

Different labelling functions might have different strengths and weaknesses. Combining their predictions into an aggregated one (e.g., an \textit{ensemble}) can assist with reducing the individual weaknesses, and obtaining a better result than individual LFs.

For this task, we compared two distinct ensemble techniques: cascade (\textbf{CSC}) and voting (\textbf{VT}). The cascade method takes the annotations from the first LF that annotated each node, and follows an ordered list of LFs: therefore, putting first the LFs with a higher JSD score, but lower annotated percentages, combines the high-quality annotation of these LFs with the ones with higher coverage, but lower quality.

On the other hand, the voting ensemble technique involves each LF casting a weighted vote for their top 10 labels. The weight of the vote is inverse to the position of the label (Table~\ref{tab:weighted_lf}). Consequently, the label with the highest probability is awarded a score of 10, while the label in the second position is given a score of 9, and so on, until the 10th label, which gets a score of 1, after they are all 0. Finally, the votes are summed, and the vector is normalized (Table~\ref{tab:voting}). Only LFs that annotate the node can cast a vote. This ensemble method removes a lot of the noise since most of the labels will have a probability of zero. 

The LFs used in these two approaches are manually selected by taking into account the individual characteristics of each LF. We use recall, percentage of unannotated nodes, the agreement between LFs, and variety to decide which LF to use in the ensemble. 

\begin{table}
\caption{Example of the VT ensemble approach for the Weka file \texttt{../classifiers/functions/SimpleLogistic.java}. }
\begin{subtable}[t]{0.4\textwidth}
\caption{Weighted predictions for the \textit{keyword}-based LF on the identifiers.}
\begin{tabular}{lcc}
\toprule
                   \textbf{Label} &  \textbf{Weight} &   \textbf{Prob} \\
\midrule
 Random Forest &      10 & 0.0561 \\
   Information Extraction &       9 & 0.0337 \\
        Anomaly Detection &       8 & 0.0337 \\
  Software Design Pattern &       7 & 0.0337 \\
Facial Recognition &       6 & 0.0337 \\
            \dots  &   \dots  & \dots \\
\bottomrule
\end{tabular}
\label{tab:weighted_lf}
\end{subtable}
\hspace{\fill}
\begin{subtable}[t]{0.4\textwidth}
\caption{Final predictions on the voting ensemble.}
\begin{tabular}{lc}
\toprule
                    \textbf{Label} &    \textbf{Prob} \\
\midrule
            Random Forest &     0.3603 \\
Logistic Regression       &     0.3603 \\
   Information Extraction &     0.3243 \\
        Linear Regression &     0.3243 \\
        Anomaly Detection &     0.2882 \\
  \dots  &    \dots \\
\bottomrule
\end{tabular}
\label{tab:voting}
\end{subtable}
\label{tab:aggregation}
\end{table}

\section{Evaluation}
\label{sec:eval}

To evaluate the quality of the annotation generated by the LFs, we use both the project-level labels assigned by the developers (i.e. the ground truth) and human evaluations of the generated labels at each level of granularity (project, packages, and files). Furthermore, we also use two automatic metrics (polarity and agreement) to get a general view of the characteristics of each LF.

\rev{Lastly, we also present the results for a baseline approach (\textbf{Rand}). We picked the random baseline as it is the only approach that does not require other data or manual annotation to generate. The random baseline consists of sampling a label from a uniform distribution over all labels.}

\subsection{Annotators Instruction}
For the manual evaluation, we used a total of 6 annotators, with four being PhD students, and two industry developers. Their background varies, with the majority having a software engineering background, while there are a couple with a more machine learning and natural language processing background.

The annotators were instructed to familiarize themselves with the project by checking the GitHub page for the project, the website, or documentation. For the project-level annotations, assign 1 whether they thought that the predicted label was correct for the corresponding project; 0 otherwise. For the package-level and file-level, since for each package/file, there were three predictions, we asked the annotators to assign 1, 2, or 3 based on which of the labels was the correct one. They were instructed to use the package/file name as a way to reduce the complexity and time required by reading the file's content. If they do not think that any of the labels are correct, then they could assign 0.

\color{black}

\subsection{Project-level Annotations}
\label{sec:eval-proj-level-ann}

The evaluation of the labelling functions was initially performed at the project-level, since we already have access to the ground truth data (i.e., the project labels available on GitHub Topics): this enabled us to estimate, at the project level, the overall performance of each LF. It has to be mentioned that the ground truth data available at the project-level is imperfect, as it contains noise (e.g., irrelevant labels) and incomplete annotations. Therefore, instead of precision, we focused on evaluating the recall measurement, which indicates how well the LFs capture the developer-assigned labels. We evaluated the recall@k, with k = 3, 5 and 10 labels.

As explained above, and in order to better understand how the LFs perform, we used the Jensen-Shannon distance (JSD) of the predictions against the uniform distribution, which allowed us to evaluate the confidence of the project-level predictions. A higher JSD value indicates that the annotations are less noisy (i.e., the peaks are more clearly distinguishable): this, when combined with recall, provides a general idea of the effectiveness of the labelling process.

Lastly, in order to evaluate the LFs' ability to capture \textit{new} application domains for the projects, we manually assessed a sample of new labels for 100 projects. The projects selected for this assessment were chosen based on their popularity, as this can increase the annotators' familiarity or reduce the time required to get familiar with the project. \revision{We pick the top-10 recommended labels for each project} and discard the ones that matched those already assigned by developers, resulting in 817 pairs (project, new label) being evaluated.

This evaluation is only performed on the best method, the voting ensemble. We utilize Cohen's kappa~\cite{landis1977agreement}, a widely used measure for intra-rater reliability. 

\subsection{Package-level Annotations}
\label{sec:eval-package-level-ann}

Similarly to the project-level annotations, we used the JSD algorithm to measure how confident the LFs are in their annotation at the package level. Unlike the pre-existing project-level labels, however, we could not access ground truth labels for packages. To address this limitation, we leveraged a characteristic that software packages should embody: all the source files contained within a package should be related to a specific functionality, and share a high cohesion within the package they belong to. 

In order to \revision{evaluate the package annotation}, we calculated a cohesion score to \revision{asses} how differently the files within a package were annotated (by the previous step) compared to \revision{the} others. This `label cohesion' score was computed by taking the average pairwise JSD values for all annotated files within a package. A higher score indicates that the LFs' annotations are more cohesive within the package, \revision{a lower score, on the other side, indicated that the labels assigned to the package files are different}.

Lastly, we performed a human evaluation over a randomly selected set of $1,000$ annotated packages, i.e., 10 for each of the 100 projects selected above. We presented the annotators with the top 3 labels for each package and asked them whether the correct label was available in the presented list. 

We evaluated whether the annotators could agree on any of the first three predicted labels. If, for example, the predicted labels for a package were A, B and C, the 2 annotators would get those three to choose from. If the human evaluations returned as C, C, we would consider the package correctly labelled (with the label in position 3). If there was a disagreement, like the annotators marked A, and C, a third annotator would perform a disagreement by picking the best or marking both as wrong.

As for the project-level labels, we utilized Cohen's kappa to measure the intra-rater reliability for these package-level labels (this is done before the disagreements are resolved).

\subsection{File-level Annotations}
\label{sec:eval-file-level-ann}
As for the package-level label predictions, we did not have ground truths to leverage to annotate source code files; furthermore, we do not have other information that can assist us as for the previous levels (\eg files in the package). Therefore, we only conducted a human evaluation to assess the quality of the file-level annotations. To this end, we randomly selected $1,000$ annotated source files, 10 from each of the 100 projects, and asked the human annotators to evaluate the proposed annotations. This evaluation was achieved with the same procedure as the package-level, by showing the annotators the top 3 labels for each source and asking which one is correct. As for previous levels, we utilized  Cohen's kappa to measure the intra-rater reliability.

\subsection{Labelling Function Statistics}
\label{sec:eval-metrics}
Besides measuring the performance of annotation produced by the LFs, we can also evaluate their characteristics. One measure we used to evaluate their performance is the \textit{polarity}, i.e., the number of unique labels the LF outputs. This is an indicator of the LF's ability to capture the labels' features in the documents. 

Another metric we used to evaluate LFs is based on measuring the \textit{agreement} between two LFs, i.e., the number of labels that the two LFs agree upon, using a pool of the top 10 predicted labels. We computed these metrics at the project-level.

\section{Results}
\label{sec:results}

The results obtained from our study are presented in this section, providing a detailed analysis of both the automated metric and human evaluation results across all levels: project (\ref{sec:result-project-level}), package (\ref{sec:result-package-level}) and file-level (\ref{sec:result-file-level}). Furthermore, we also present the statistics of the considered labelling functions (\ref{sec:result-stats}).

\subsection{Project-level annotations}
\label{sec:result-project-level}

We start with evaluating the project-level annotations as they allow us to assess the effectiveness of our LF automatically, making it easier to pick the best one and perform manual evaluation only on it.

The project-level evaluation will allow us to answer \textbf{RQ3}:

\begin{quote}
    \textit{\textbf{RQ3}: Does aggregation at the project-level provide an effective means for capturing the content of projects?}
\end{quote}

\color{black}

\begin{figure*}[htbp!]
    \centering
    \includegraphics[width=\textwidth]{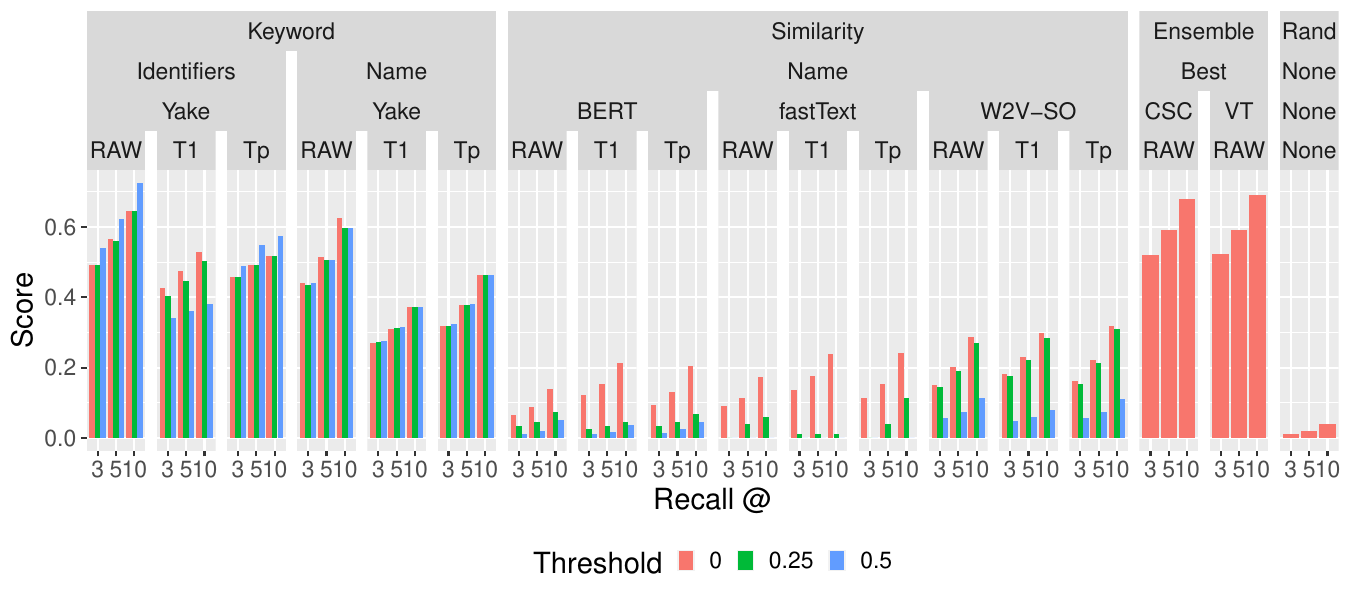}
    \caption{Recall scores for project level annotations considering different numbers of top labels.}
    \label{fig:recall}
\end{figure*}

The recall of the three labelling functions, computed at various thresholds (i.e., using 3, 5, and 10 labels) is shown in Figure~\ref{fig:recall} for the project-level annotations. 

Overall, we noticed that the LFs with the highest recall are the keyword-based ones: all the \textit{similarity}-based LFs score noticeably worse regarding their recall. The general higher recall for \textit{keyword}-based LFs is due to how they predict more labels that are not in the first (more likely) positions. For the similarity-based ones, while their performance is not optimal, they are still able to pick the best label in the first positions; therefore, a reduction of noise is more beneficial. 

Considering the \textit{keyword}-based LFs, the RAW predictions are better than the transformed ones (T1, and Tp). In contrast, when a transformation is applied, the \textit{similarity}-based LFs show an improvement for all except one case (T1 for W2V-SO).

The high noise for the \textit{similarity}-based LFs is also visible from the filtering (using the threshold at 0, 0.25 or 0.5), where their performance suffers noticeably. Filtering strongly affects the \textit{identifiers}-based LFs without any transformation, whereas filtering with a threshold of 0.5 achieves the best recall score.

Among the \textit{similarity}-based LFs, the best recall is achieved by the Word2Vec model trained on Stack Overflow (W2V-SO), which suggests that domain knowledge is needed to achieve good results. 

Lastly, when considering the ensemble, we see similar results between the two approaches (CSC and VT), with a slightly higher score for VT with 10 labels. These results align with the \textit{keyword}-based LFs with a filtering of $0.5$. 

\rev{All the methods outperform the random baseline, however, for the \textit{similarity}-base approaches, when increasing the threshold, the performance reaches the one of the random baseline.}

\begin{figure*}[htbp!]
    \centering
    \includegraphics[width=\textwidth]{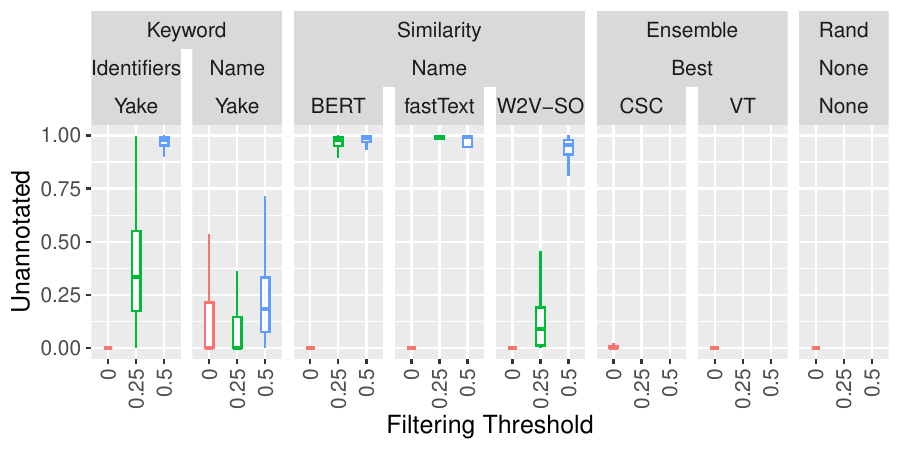}
    \caption{Distribution of the percentage of unannotated nodes.}
    \label{fig:unannotated_percent}
\end{figure*}

While having similar recall scores, using this as the only indicator to decide which LF is the best is insufficient. As mentioned earlier, increasing the filtering threshold might remove some nodes based on the annotations' noise. Therefore, we also need to consider the number of nodes that are not being labelled. Figure~\ref{fig:unannotated_percent} shows us the percentage of unannotated nodes for each LF. As clearly visible, using a threshold of $0.5$ negatively affects the amount of annotated nodes. In most cases, the number of unannotated reaches $90\%$, except the \textit{name}-based LF, where it only reaches $50\%$. A threshold of $0.25$ also negatively affects the \textit{similarity}-based LFs, but not the keyword-based ones, indicating higher confidence in the predictions for the \textit{keyword}-based LFs.

\begin{figure*}[htbp!]
    \centering
    \includegraphics[width=\textwidth]{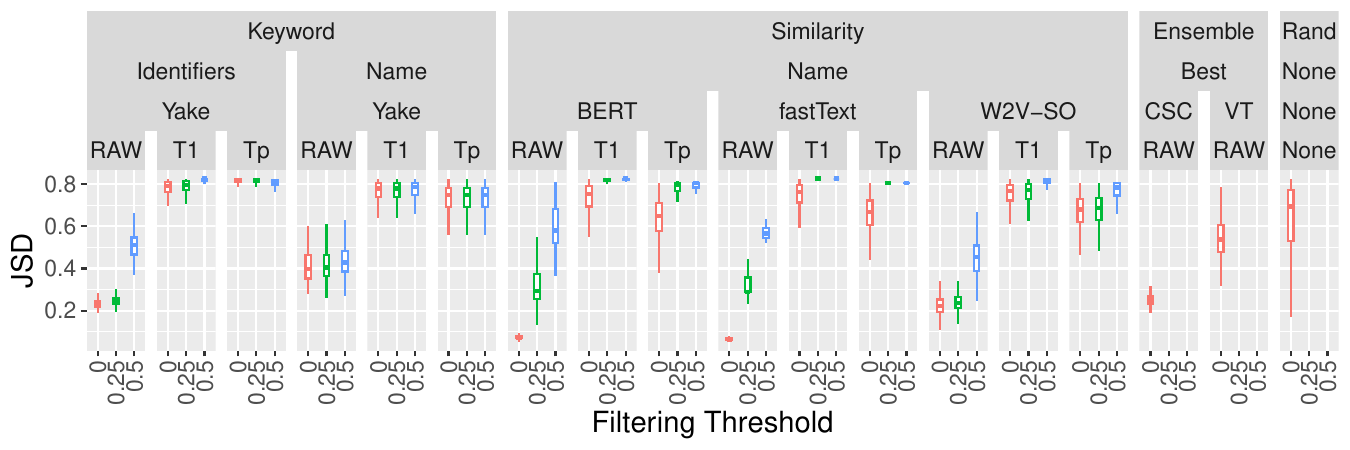}
    \caption{Project-level JSD distribution for the different LFs.}
    \label{fig:project_jsd}
\end{figure*}

We can measure this confidence with the JSD of the prediction against a uniform distribution (highest entropy), as shown in Figure~\ref{fig:project_jsd}. In most cases, a high filtering threshold is beneficial; however, it is not ideal given the large number of unannotated nodes. A more conservative threshold of $0.25$ slightly increases the JSD and does not significantly affect the amount of annotated nodes. Overall, the filtering, at the project level, while improving the recall, does not seem to be a crucial aspect, given the downside of fewer annotated nodes.

\revision{We can see that the random baseline has a very high variance in the JSD, indicating that the labels are all over the place, however, given the fact that there is only one label for each file, the score is high.}

\revision{Focusing on the ensemble, we can notice that for the recall, it performs similarly to the \textit{keyword}-based LFs. However, if we also consider the number of unannotated nodes, we can see that, with the \textit{ensemble}-based LFs, we achieve a near zero percentage, while the \textit{keyword}-based ones present more noise. This higher noise is also captured by the lower JSD. Therefore, considering all these metrics, we pick the voting ensemble (VT) as the best LF for human evaluation.}

\newpage

Therefore, we can answer our \textbf{RQ3}:

\begin{mybox}
The voting ensemble LF achieves a recall for the developer-assigned labels between 50\% (recall@3) and 70\% (recall@10). This shows the effectiveness of the LF in capturing the pieces of information at the file-level, and that the signal is strong enough not to get suppressed by the aggregation.    
\end{mybox}


Now that we measured the ability to discover the developers' assigned labels, we are interested in answering \textbf{RQ4}:

\begin{quote}
    \textit{\textbf{RQ4}: Can file-level annotations allow the discovery of new topics within projects?}
\end{quote}

\begin{figure}[htb!]
    \centering
    \begin{subfigure}[b]{0.29\textwidth}
         \centering
         \includegraphics[width=\textwidth]{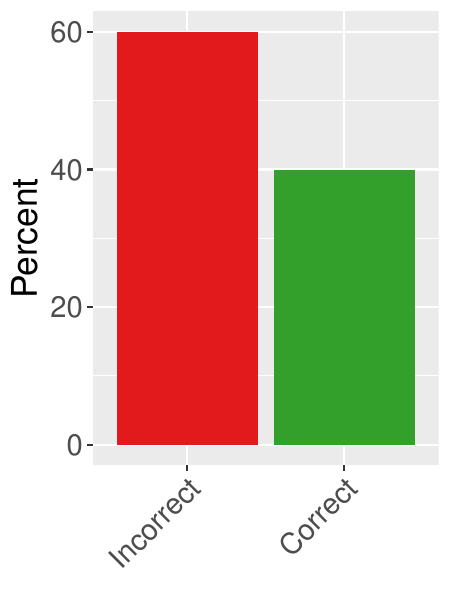}
         \caption{Percent of correct and incorrect new labels.}
         \label{fig:human_eval_project}
     \end{subfigure}
     \hfill
     \begin{subfigure}[b]{0.68\textwidth}
         \centering
         \includegraphics[width=\textwidth]{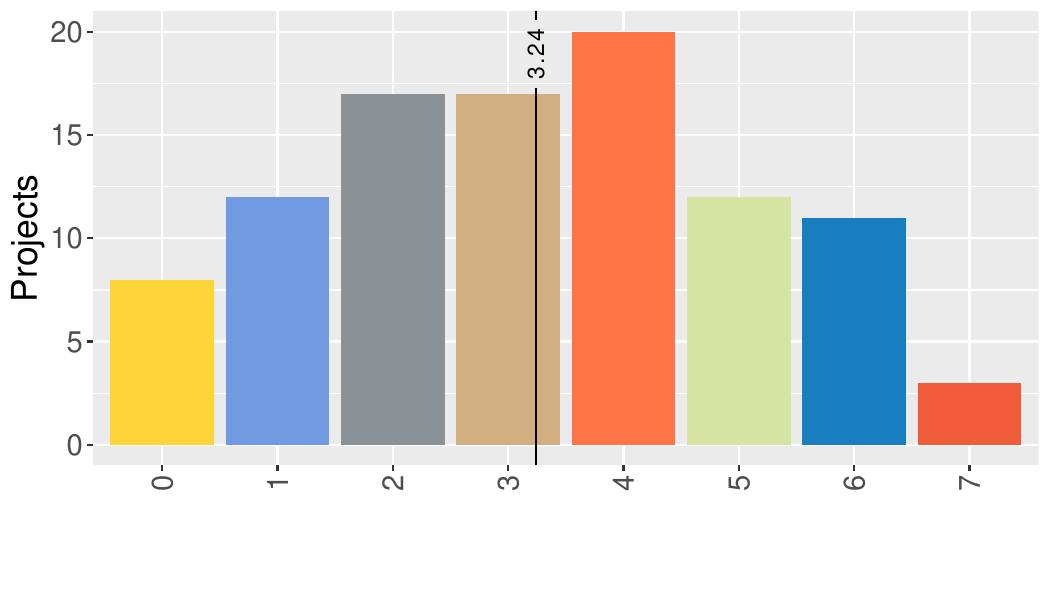}
         \caption{Distribution of the number of new topics discovered. The mean is shown with a vertical line.}
         \label{fig:new_topics}
     \end{subfigure}
    \caption{Results of the human evaluation at the project-level.}
    \label{fig:project_human_eval}
\end{figure}

Using human annotators, we evaluated the ability of the models to find new topics for the project. Figure~\ref{fig:project_human_eval} presents the human evaluation results on the newly discovered topics. We first measure the intra-rater agreement, shown in Table~\ref{tab:agreement}. We can see a moderate agreement~\cite{landis1977agreement} of $0.55$ between the annotators, with 21\% of the labels requiring resolution of the disagreement. The disagreement was resolved using a third annotator before computing the metrics. Figure~\ref{fig:human_eval_project} shows that 40\% of the topics identified are correct, while the remaining 60\% are not. The number of newly identified topics varies across the projects (Figure~\ref{fig:new_topics}), with, on average, three new topics being found for each project, and in a couple of cases, we reach seven new topics.   

\begin{table}[htbp!]
\centering
\caption{Cohen's Kappa for the intra-rater agreement at the various levels, and the percentage of examples annotators disagree on.}
\begin{tabular}{lcc}
\toprule
\textbf{Level}   & \textbf{Kappa} & \textbf{\% Disagreement} \\
\midrule
Project & 0.55  & 21\%            \\
Package & 0.46  & 35\%            \\
File    & 0.50  & 32\%             \\
\bottomrule
\end{tabular}
\label{tab:agreement}
\end{table}

These results show the ability of our approach to not only find the developer-assigned topics, as we saw using the recall (Figure~\ref{fig:recall}), but also find new relevant topics for the project using the file-level information.

An example of newly predicted project-level labels can be seen in Table~\ref{tab:proj-prediction}. While the examples are above average regarding the number of new topics identified, they give us an idea of what the predictions look like. For Weka, we can see that while the wrongly predicted labels are not relevant, we can argue that both \lbl{Data Structure} and \lbl{Database} can be present. Similarly, for Pumpernickel, we have topics that, while incorrect, are closely related to the domain of the application (\eg \lbl{Animation}).  

\begin{table}[htbp!]
\centering
\caption{Top predicted project-level labels  with the human evaluation for Weka and Pumpernickel.}
\begin{tabular}{lc|lc}
\toprule
\multicolumn{2}{c|}{\textbf{Weka}}              & \multicolumn{2}{c}{\textbf{Pumpernickel}}      \\ \midrule
\textbf{Predicted}                & \textbf{Eval} & \textbf{Predicted}               & \textbf{Eval} \\ \midrule
Semi-supervised Learning  & 1          & Image Editing            & 1          \\
Database                  & 0          & Image                    & 1          \\
Data Structure            & 0          & Text Editor              & 0          \\
Naive Bayes Classifier    & 1          & Image Captioning         & 1          \\
Big Data                  & 1          & Digital Image Processing & 1          \\
Data Binding              & 0          & Animation                & 0          \\
Logistic Regression       & 1          & GUI                      & 1          \\
Data                      & 1          & Text Processing          & 1          \\
User Interface            & 1          & Web Browser              & 0          \\ \midrule
\end{tabular}
\label{tab:proj-prediction}
\end{table}


We can summarize the findings and answer \textbf{RQ4}:

\begin{mybox}
File-level annotations can play a crucial role in discovering new project topics. The results suggest that around 40\% of new predictions are correct. It is estimated that three new topics (besides those already set by the originating developers) can be discovered for every project, on average. 
\end{mybox}


\subsection{Package-level Annotations}
\label{sec:result-package-level}

We are now moving to the evaluation of the annotations of packages, which will allow us to answer \textbf{RQ2}:

\begin{quote}
    \textit{\textbf{RQ2}: Does aggregation at the package-level provide an effective means for capturing the content of packages?}
\end{quote}

\begin{figure*}[htbp!]
    \centering
    \includegraphics[width=\textwidth]{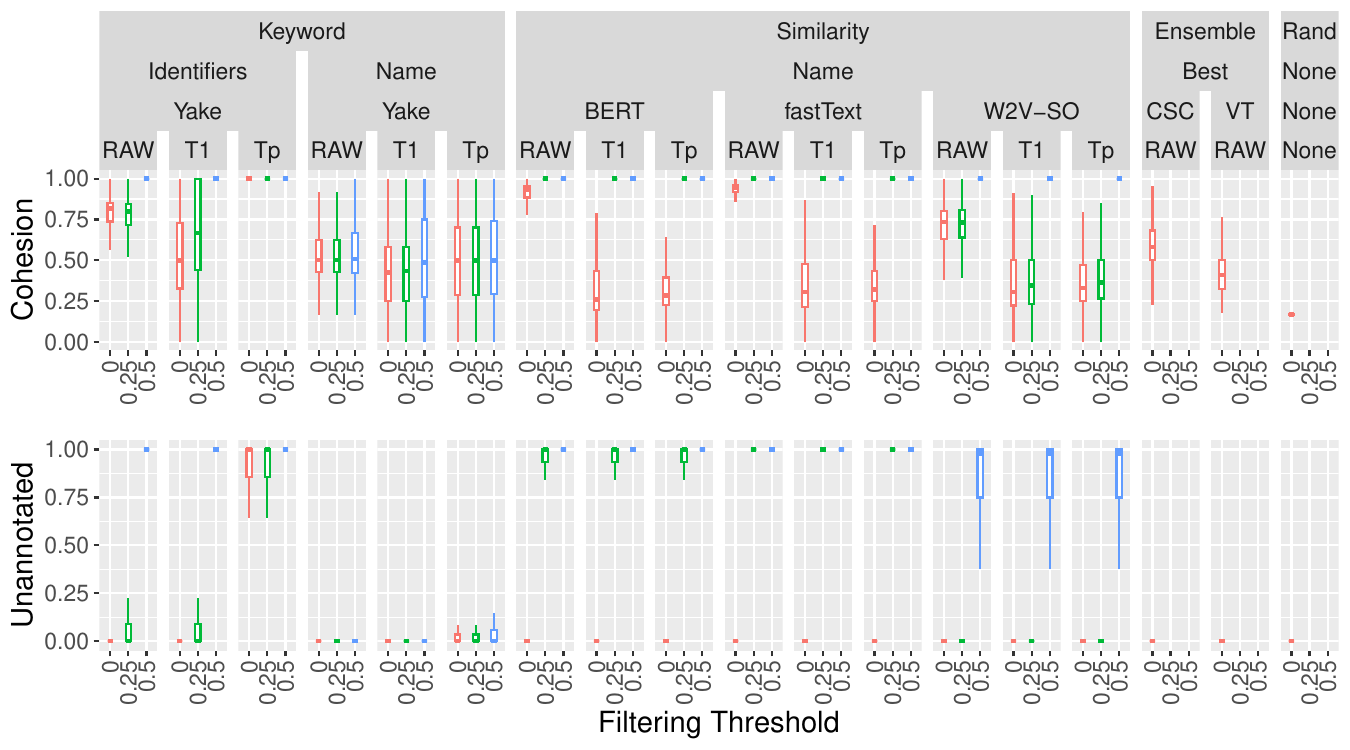}
    \caption{Package stats. Cohesion, average pairwise JSD among files in the same package. And percentage of unannotated packages.}
    \label{fig:package_stats}
\end{figure*}

Given the inability to use ground truths, we focus on the cohesion of the annotation in the package. Figure~\ref{fig:package_stats} shows that the variance is high in most LFs; however, the average scores are also high. \rev{The random baseline scores are around 0.20, indicating almost no cohesion, as expected when assigning labels in a random fashion. Moving to the \textit{keyword}-based LFs,} we can see that they perform on average better \rev{than all the others}. In particular, \textit{name}-based LF has an average cohesion of $0.5$; in contrast, the \textit{identifiers}-based LF performs much better, averaging around $0.8$. On the similarity side, the best case is the W2V-SO without filtering, with the other performing below $0.5$ on average. 
Lastly, the cohesion for the filtering threshold of $0.5$ is almost always at $1$; however, in all cases, it is due to the high amount of unannotated packages. 

The number of unannotated packages follows an interesting pattern. In all cases except for the W2V-SO, we have either a very low percentage of unannotated packages or almost all unannotated. 
Again, high thresholds are not optimal, but compared to the project-level, a moderate one is ideal for slightly better cohesion and little effect on the number of unannotated packages.

Moving to the ensemble approaches, while there is a higher cohesion for the cascade approach, the difference with the voting method is not substantial enough to make the cascade approach a better strategy when considering the much higher difference in JSD at the project-level.

Moving to the human evaluation, we have a decrease in the disagreement (Table~\ref{tab:agreement}) compared to the project-level; however, with a kappa of $0.46$ we can consider it still moderate agreement. The package level has the highest level of disagreement, with 35\% of the considered samples requiring resolution from a third annotator. 

After resolving disagreements, we can see in Figure~\ref{fig:human_eval_results} that most predictions are correct for the package-level annotations, with only 43\% of examples being incorrectly labelled. Moving to the correct instances, most of the accurate labels are in the first position, with 34\% of cases having the correct label in the first position. The aggregation is also effective at capturing the semantics of a package.  An example of the predictions and the human evaluation is presented in Table~\ref{tab:package-predictions}.

\begin{table}[htb!]
\centering
\caption{Example of 5 Weka packages with the human-assigned evaluations and their position in the sorted prediction list.}
\begin{tabular}{llc}
\toprule
\textbf{\textbf{Package}}                                   & \textbf{Prediction}          & \multicolumn{1}{l}{\textbf{Pos}} \\ \midrule
classifiers.evaluation.output.prediction  & Data                & 1                            \\
classifiers.timeseries.core               & Machine Learning    & 3                            \\
datagenerators.classifiers.regression     & Logistic Regression & 1                            \\
core.matrix                               & -                   & 0                            \\
datagenerators.classifiers.classification & Machine Learning    & 3                            \\ \bottomrule
\end{tabular}
\label{tab:package-predictions}
\end{table}

One aspect to consider for the results is that when evaluating, we assume that the package content has high semantic cohesion, which is not always the case. The lack of cohesion also affects the annotation results at the package level.

Summarizing the results, we can answer \textbf{RQ2}:
\begin{mybox}
Aggregation at the package-level can indeed provide an effective means for capturing the content of packages. Furthermore, the results suggest that combining the file-level annotations obtained from the voting ensemble LF (VT) makes it possible to accurately annotate at least 50\% of the examples at the package level.
\end{mybox}

\subsection{File-level Annotations}
\label{sec:result-file-level}

In this section we are going to evaluate the file-level annotation, which will allow us to answer our \textbf{RQ1}:

\begin{quote}
    \textit{\textbf{RQ1}: To what extent is weak labelling effective in capturing the semantic content of files for annotation purposes?}
\end{quote}

\color{black}

At the file-level, we already had a view of the number of unannotated nodes in Figure~\ref{fig:unannotated_percent}. 
Therefore, this section will only present the human evaluation results. 

\begin{figure*}[htbp!]
    \centering
    \includegraphics[width=0.87\textwidth]{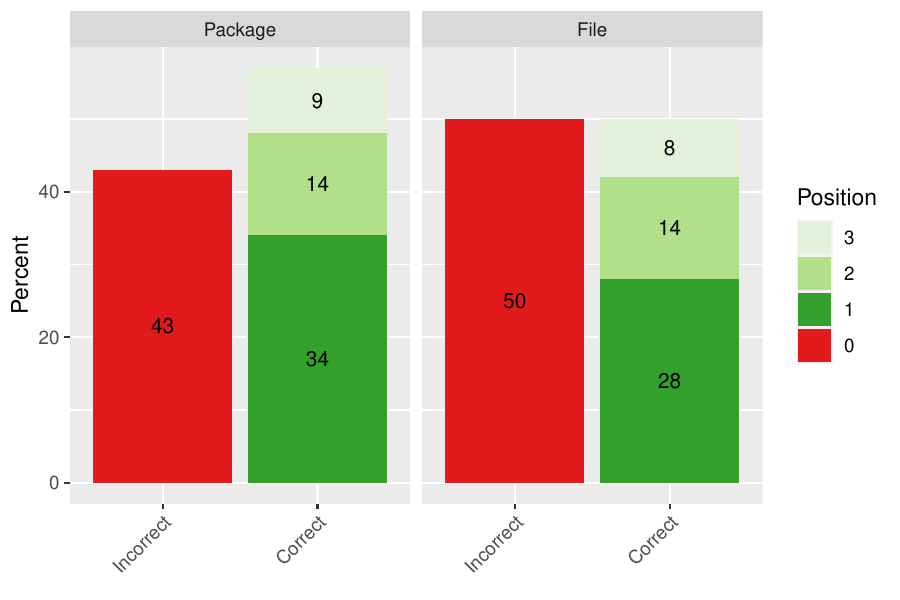}
    \caption{Results of human evaluation at the package and file-level. Percentage of incorrect and correct labels, the correct labels are separated by their position in the prediction.}
    \label{fig:human_eval_results}
\end{figure*}

In Figure~\ref{fig:human_eval_results}, along with the package results, we can also see the file-level results. In this case, the percentage of incorrectly labelled files is higher, reaching 50\%. However, the most likely prediction, in position 1, is accurate in most correct cases.

A qualitative view of the file-level annotations can be seen in Table~\ref{tab:file-predictions}. In this subset, most of the examples are correct; however, when examining the specific case of the \texttt{\dots/gui/beans/Note.java} file, given the file path, it is easy to say that it should be labelled as \lbl{UI}. This raises the idea that using extra information from the package might benefit the approach. 

Lastly, we should also consider that some files might not have enough information for proper classification or contain a mix of topics in a single file. For example, while the file \texttt{gui/beans/AbstractTestSetProducerBeanInfo.java} has been correctly labelled, its label is in position two. However, the best prediction, in position one, is \lbl{Machine Learning}. As the filename suggests, there is a mix of ML and GUI terms; however, the file is responsible for \textit{UI} for a \textit{ML} application. Therefore, it is also important to understand that some files might be misclassified due to this overlap or cases where a file is concerned with more than one responsibility.

\begin{table}[htbp!]
\centering
\caption{Example of 5 Weka files with the human-assigned predictions and their position in the sorted prediction list.}
\begin{tabular}{llc}
\toprule
\textbf{File}                                   & \textbf{Prediction}       & \multicolumn{1}{l}{\textbf{Pos}} \\ \midrule
\dots/clusterers/NumberOfClustersRequestable.java    & Machine Learning          & 2                            \\
\dots/gui/beans/AbstractTestSetProducerBeanInfo.java & User Interface            & 2                            \\
\dots/pmml/jaxbbindings/BoundaryValueMeans.java      & Data                      & 2                            \\
\dots/gui/beans/Note.java                            & -                         & 0                            \\
\dots/classifiers/functions/SimpleLogistic.java      & Logistic Regression       & 1                            \\ \bottomrule
\end{tabular}
\label{tab:file-predictions}
\end{table}

In conclusion, the effectiveness of weak labelling in capturing the semantic content of files for annotation purposes can be evaluated based on the positive rate achieved during manual evaluation. In the case mentioned, the weak labelling approach captured the semantic content of the files with a positive rate of 50\%. Furthermore, the proposed approach has a near-zero amount of unannotated files, while maintaining a high JSD.

Finally, we can summarize the results and answer \textbf{RQ1}: 
\begin{mybox}
Our weak labelling approach achieved a 50\% positive rate in capturing the semantic content of the files during manual evaluation. This indicates a moderate level of effectiveness and has the potential to be useful for annotation purposes, albeit with certain limitations.
\end{mybox}

\subsection{Labelling Function Statistics}
\label{sec:result-stats}
We can use the \textit{polarity} and the \textit{agreement} measurements to understand better the used LFs' behaviour. 

\begin{figure}[htbp!]
    \centering
    \includegraphics[width=.8\textwidth]{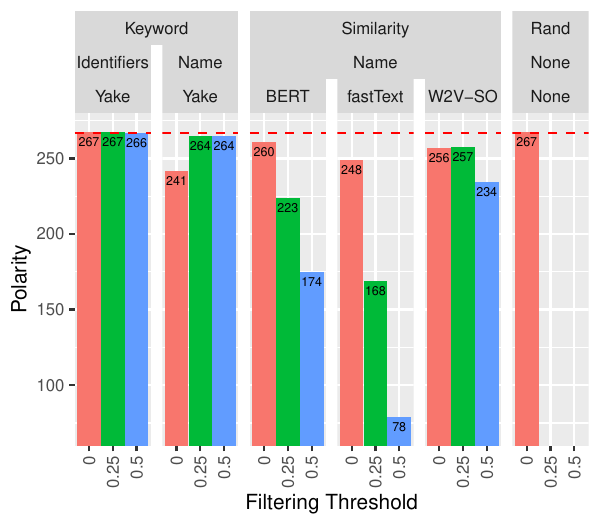}
    \caption{Set of unique top 10 labels each LF outputs with the raw annotations at the project-level. The dashed line is the number of total labels (267).}
    \label{fig:polarity}
\end{figure}

\paragraph{Polarity -- } Figure~\ref{fig:polarity} presents the polarity score: we can see that the \textit{keyword}-based approaches produce better results than the \textit{similarity}-based approaches. The \textit{identifiers}-based LF can return almost all the labels at all levels of filtering, indicating its ability to identify all the classes. For the \textit{name}-based LF, we noticed something interesting: the unfiltered annotations (threshold = 0) have the lowest amount of labels predicted, and that can be a symptom of bias for specific labels in the more uncertain cases (\ie label A has a slightly higher similarity in the majority of cases when all other labels are low as well).

For the \textit{similarity}-based LFs, we noticed a significant decrease in the polarity with an increasing threshold for all functions. Furthermore, when using a \textit{similarity}-based LFs, no LF can predict all the labels independently of whether there is a filtering of uncertain nodes. 

Lastly, for the \textit{name}-based LF, we noticed an increase in the polarity when the filtering threshold increased, in contrast to the other LFs. This can be due to some labels that are favoured when uncertain, similar to the \textit{similarity}-based LFs; however, in this case, it can be due to the lack of significant keywords in the file, and one label having a general one that will boost its probability.

\begin{figure}[htbp!]
    \centering
    \includegraphics[width=.85\textwidth]{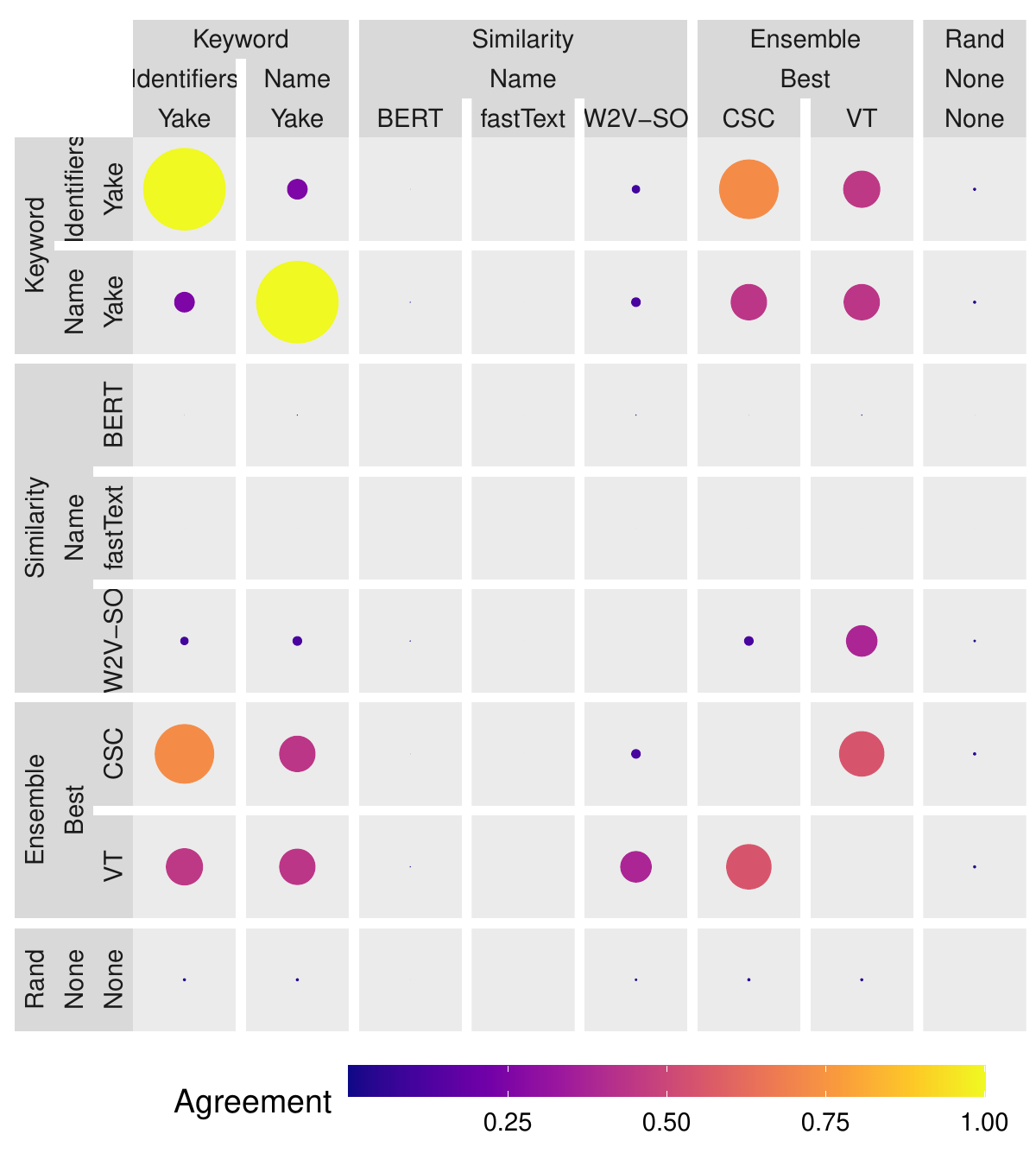}
    \caption{Agreement in annotation between LFs.}
    \label{fig:lf_agreement}
\end{figure}

\paragraph{Agreement -- } To check the similarities in annotation behaviour between the LFs, we also evaluated the \textit{agreement} metric: Figure~\ref{fig:lf_agreement} shows the ratio of labels predicted between the LFs. 

The scores show a minimal overlap between the LFs, excluding the ensemble methods. Albeit low, this overlap can boost the better labels, which is the idea behind using the ensemble method. 
The \textit{similarity}-based LFs have the slightest overlap, while the keyword-based approaches share a higher agreement. Concerning the ensemble, we notice that the CSC approach is biased towards the first LF, the identifiers-based one in our case. In contrast, the VT approach considers all LFs predictions equally (note that BERT and fastText were not used in the ensembles due to their low recall and high noise). \revision{Lastly, another reason why we chose the VT ensemble can be seen by checking the agreement between the LFs, and the ensemble methods. As respected in the cascade, most agreement is found in the first LF in the cascade, the \textit{identifier}-based LF. However, we can see that the agreement between the \textit{keyword}-based LF, and the W2V-SO overall is minimal, indicating that they identify different labels. This difference and the reasonable results in terms of recall for the W2V-SO LF, suggest that using this information can be beneficial. The VT ensemble has a higher agreement, indicating that the labels are considered. While there is no gain in recall, it might help discover new labels.} \rev{Lastly, as expected, the random baseline has almost no intersection with the other approaches.}

\section{Discussion}
\label{sec:discussion}
In this section, we will discuss qualitative stances on the proposed approach.

\revision{The results showed the effectiveness of automating the annotation of files in software repositories. However,} an important aspect to report, and that can be noticed in both Tables~\ref{tab:package-predictions} and \ref{tab:file-predictions}, is the lack of specificity exhibited by some assigned labels. One such example is the \revision{Weka source code file:}\texttt{\dots/clusterers/NumberOfClustersRequestable}, for which a more suitable label would be \lbl{Clustering}. 
This is also noticeable in other instances of the evaluated examples: this phenomenon is likely because more examples feature general labels, making them more probable since they contain a broader list of keyword terms. Building a taxonomy incorporating explicit hyponymy and hypernymy relationships\footnote{In natural language, a hypernym describes a broader term, whereas a hyponym is a more specialised word. For example, `Deep Learning' is the hypernym, while `Convolutional Neural Network (or `CNN') is the hyponym.} within the labels would address this issue.

\revision{Another aspect we can notice from the results is the significant difference between the \textit{keyword}-based LFs, and the \textit{similarity}-based ones. One reason we can point to this large difference is the domain. The \textit{keyword}-based approaches use domain information, while the \textit{similarity}-based ones are more general approaches, except for the W2V-SO, which also performs better. Finding better models that encode domain knowledge can help the results for the \textit{similarity}-based LFs.}

Furthermore, while not part of our RQs, the filtering and transformations of the predictions were important aspects evaluated in this work. While the final method mostly uses raw predictions, filtering and transformations can be helpful in some cases, especially with very large code bases. As seen from the previous metric, filtering, by removing noisy annotations, can preserve (fewer) high-confidence annotations. Therefore, an optimal filtering threshold allows only confident annotation; however, the downside of reducing the amount of annotated nodes for very large-scale datasets can be less of an issue. Furthermore, the transformation of the annotations aids with removing some noise while preserving a good amount of information.

\rev{Moreover, one issue with our approach is that the average gives more weight to frequent topics in files, not central to the software's functionality. Therefore, a software application domain might end up being a secondary, or lower, label while backend functionality might get the main label spot. This issue can be addressed in future work by incorporating the structural information from the dependency graph. }

\revision{Lastly, currently, we are using a na\"ive ensemble approach for the combination of the labels; however, this does not take into account that some LFs might have better performance on different labels, which will also allow the use of per domain (label) specific language models (e.g., a biology LM for the biology labels, and a finance one for finance labels). Further research could explore this direction.}

\section{Uses}
\label{sec:uses}

From a practitioner's point of view, multi-granular annotations can be leveraged to automatically generate semantically labelled graphs that depict a software system's internal semantic content, as seen in Figure~\ref{fig:package_labels}. This can significantly reduce the time spent on software comprehension, which typically accounts for around 58\% of the development time~\cite{xia2019measuring}. This is particularly beneficial in industry settings, where newly recruited developers must be trained to understand the business process and relevant code, which can be time-consuming.

Additionally, this information can assist with automatic documentation generation, similar to Software Architecture Reconstruction (SAR) in microservices architectures\cite{Rademacher2020SAR, walker2021automatic}. This has significant practical implications as it can help to facilitate the retrieval of components from open-source platforms like GitHub, promote software reuse, and improve overall development efficiency.

From a research perspective, multi-granular annotations can be employed to investigate context-driven research in the software engineering domain. This approach is consistent with recent studies emphasising the importance of considering the context in software engineering studies~\cite{briand2012embracing, briand2017case}.

\section{Threats to Validity}
\label{sec:threats}

We will present the \emph{construct validity}, \emph{internal validity}, and \emph{external validity} that we encountered during our study, and we discuss how we addressed them. 

\subsection{Construct Validity}

The recall is used to measure the quality of the file-level annotations indirectly; therefore, we do not have a direct measure of the quality of the file and package-level annotations. We mitigate this issue by implementing human validation on a representative subset of examples from both files and packages. By doing so, we can obtain a more accurate evaluation of the quality of annotations at these levels.

The JSD is a valuable metric for measuring the noise in assigned labels, although it is not necessarily an indicator of quality. However, high JSD values (i.e. low noise) can indicate that the LFs generate fewer high-likelihood predictions. This desirable behaviour suggests that the LFs provide fewer and more specific candidates. As a result, even though the JSD score may implicitly favour the voting ensemble, this behaviour ultimately leads to prediction with very few candidates and nearly no noise. 

\subsection{Internal Validity}
Analysing the labelling is inherently subjective since it involves natural language and requires prior knowledge of various application domains. Moreover, manual evaluations were conducted solely based on the names, which can make it more challenging due to the limited information available. However, we have mitigated this potential issue by ensuring that two annotators evaluate each example and a third annotator resolves any discrepancies.

\subsection{External Validity}

We obtained a comprehensive list of terms for our labelling functions by extracting keywords from a large pool of projects. However, it's worth noting that the keywords were only taken from a sample of Java projects, which may limit their generalizability. One approach to address this limitation is expanding the project pool to include more programming languages. Fortunately, language-specific parsers can be used to quickly adapt our approach to different programming languages, which can help improve the LFs' generalizability.

\section{Related Work}
\label{sec:sota}

In the context of our research, we have identified two closely related areas of study: software classification and similarity, and program comprehension. This section comprehensively reviews the relevant prior work in these areas, highlighting their contributions and limitations.

\subsection{Software Classification and Similarity}
One of the initial works on software categorization is MUDABlue~\cite{Kawaguchi2006MUDABlue}, which applies information retrieval techniques to categorize software into six SourceForge categories. In particular, they use Latent Semantic Analysis (LSA) on the source code identifiers of 41 projects written in C. 

Following MUDABlue, Tian et al. propose LACT~\cite{tian2009lact}, an approach based on Latent Dirichlet Allocation (LDA), a generative probabilistic model that retrieves topics from text datasets, to categorize software from identifiers and comments in source code. In addition, they use a heuristic to cluster similar software. 

Altarawy et al. expand LACT into LASCAD~\cite{altarawy2018lascad}, by replacing the heuristic in LACT with hierarchical clustering using cosine similarity over the LDA vectors.

Another approach that uses topic modelling is proposed by Sharma et al.~\cite{sharma2017cataloging}, using a combination of topic modelling and genetic algorithms called LDA-GA~\cite{panichella2013ldaga}. They apply LDA topic modelling on the \README files, and optimize the hyperparameters using genetic algorithms. While LDA is an unsupervised solution, humans are needed to label the topics from the identified keywords.

A different approach was adopted in~\cite{vasquez2014api}; they take API packages, classes, and methods names and extract the words using the naming conventions. Following~\cite{Ugurel2002classification}, they use information gain to select the best attributes for the classification and then apply different machine learning methods. 

CLAN~\cite{mcmillan2012clan} provides a way to detect similar apps based on the idea that similar apps share some semantic anchors. Given a set of applications, they create two terms-document matrices, one for the structural information using the package and API calls, the other for textual information using the class and API calls. Both matrices are reduced using LSA, and the similarity across all applications is computed.  Lastly, they combine the similarities from the packages and classes by summing the entries. In \cite{linares2016clandroid}, they propose CLANdroid, a CLAN adaptation to the Android apps domain. 

Nguyen et al.~\cite{nguyen2018crosssim} propose CrossSim, an approach that uses the manifest file, project files, and the list of contributors of GitHub Java projects to create an RDF graph. Projects and developers are nodes, and edges represent the use of a project by another or that a developer is contributing to that project. They use SimRank~\cite{glen2002simrank} to identify similar nodes in the graph. According to SimRank, two objects are considered similar if similar objects reference them.

Recent research has focused on utilizing GitHub as a primary source for classification. In~\cite{sipio2020naive}, a multi-label classifier is proposed to predict a curated list of topics based on the \README of a GitHub repository. The content of the \README files is encoded using the TF-IDF weighting scheme as a preprocessing step. A probabilistic model called Multinomial Na\"{i}ve Bayesian Network (MNB) is then utilized to recommend new potential topics for the project. This work has been extended with the development of TopFilter~\cite{rocco2020topfilter}, which combines the MNB network with a collaborative filtering engine to incorporate non-featured topics in the recommendation list. The system represents repositories and topics in a graph-based structure, and the underlying recommendation algorithm computes cosine similarity using featured vectors to suggest the most similar topics. \revision{Moreover, the authors extended TopFilter, and proposed HybridRec~\cite{rocco2023hybridrec}, which deals with the issues of unbalanced data by using a combination of stochastic and collaborative filtering recommendation strategies. The stochastic part uses Complement Na\"ive Bayesian Network, similar to TopFilter. The Collaborative part encodes the projects' topics and looks for projects with similar topics. The final recommendation is a joined list of topics.}

Several approaches have been proposed that utilize neural networks for code classification. LeClair et al.~\cite{leclair2018neural} and Ohashi et al.~\cite{ohashi2019cnn_code} both use convolutional neural networks (CNN) as their model. LeClair et al. use a C-LSTM, a combination of CNN and recurrent neural networks, with the project name, function name, and function content as input. Ohashi et al. use a binary matrix representation of C++ keywords and operators to classify short, single-file programs into six computer science and engineering categories. 

Another approach utilizing a CNN as its classifier is HiGitClass~\cite{zhang2019HiGitClass}. HiGitClass uses a heterogenous graph to model the co-occurrence of multimodal signals in a repository, including user, repository name, topics (labels), and README. They use a topic modelling approach to learn word distribution and generate documents to train a CNN for classification. The word embeddings are created using ESIM~\cite{shang2016esim}, a meta-path guided heterogeneous network embedding.

Taking a more NLP-inspired approach, based on the distributional hypothesis: `\textit{A word is characterized by the company it keeps}'~\cite{firth1957studies}, \cite{theeten2019import2vec} proposes a neural network-based approach for generating embeddings of libraries by leveraging import statement co-occurrences. Their method involves training a semantic space that captures the proximity between libraries that appear together in a given context. While they don't directly perform any classification, the resulting embeddings can be utilized for measuring similarity or training classification models.

With the popularity of large language models like BERT~\cite{devlin-etal-2019-bert} in various domains, in Repologue~\cite{izadi2020topic}, they exploit its ability in the software classification task. They use a multimodal approach that uses project names, descriptions, READMEs, wiki pages, and file names concatenated together as input to BERT. Then, they apply a fully connected neural network to predict multiple labels. Their dataset has also been used in~\cite{Widyasari2023extreme}, to evaluate the performance of extreme multi-label~\cite{wei2022xml} classification models. \revision{Furthermore, the authors expanded their work~\cite{izadi2023sedgraph} by builing a custom knowledge graph (SED-KGraph) with extra information like whether the label is a `\textit{field}', a `\textit{event}', a `\textit{programming-language}'. Lastly, they apply a recommender system for the suggestion of the topics.}

Similarly, GHTRec~\cite{9590294} has been proposed to recommend personalized trending repositories, \ie a list of most starred repositories, relying on the BERT language model and GitHub Topics. Given a repository, the system predicts the list of topics using the preprocessed \README content. Afterwards, GHTRec infers the user's topic preferences from the historical data, \ie commits. The tool eventually suggests the most similar trending repositories by computing the similarity on the topic vectors, \ie cosine similarity and shared similarity between the developer and a trending repository. 

Another multimodal approach has been proposed in Repo2Vec~\cite{omar2021repo2vec}. They use a concatenation of three embeddings created respectively from the repository metadata (title, topics, description, and \README), the tree structure of the directory, and the source code. For the metadata, they use~\cite{doc2vec}, while for the directory structure, they use node2vec~\cite{node2vec}. The source code embedding is obtained using the approach proposed in~\cite{compton2020embedding}, which uses code2vec~\cite{uri2019code2vec} on each method in each file, and they aggregate the embedding using the mean up to the repository level. 

In a similar concept to CrossSim, in~\cite{qian2022rep2vec}, they build a heterogeneous graph of various repositories, developers, and topics from GitHub and perform repository embedding. Each node in the graph is then represented using an embedding. Topics are embedded using BERT; developers are a combination of the metadata of their repositories and their profile data. The repository node is obtained by embedding source code and metadata using BERT. Lastly, the graph and features are used as input to a graph neural network that will refine the embeddings by using the information of neighbouring nodes.

\subsection{Program Comprehension}

Various approaches focus on assisting developers with program comprehension; while we focus on classification, there are intersections between the two research areas. We, therefore, present the relevant work in this section. 

One initial approach is proposed by Kuhn et al.~\cite{kuhn2007semantic}. They propose an unsupervised approach for extracting terms from source code files. Furthermore, they perform clustering using LSA and Singular Value Decomposition. However, this approach requires human annotation for the identified topics in the code.

Various approaches similar to Kuhn et al. have been proposed. For example, TopicXP~\cite {savage2010topicxp} is an approach to identify topics in source code based on LDA instead of LSA. In~\cite{Ieva2019deploying}, they propose a method for software comprehension on large code bases that uses keywords, similar to~\cite{kuhn2007semantic}. They also add structural information from the call graph for various analyses, including feature location, semantic clone detection, summary generation, and more. Lastly, in ~\cite{sun2017clustering}, they use LDA to extract topics to annotate packages to assist developers during software maintenance and evolution. 
As in Kuhn et al.~\cite{kuhn2007semantic}, these approaches do not perform classification.

\section{Conclusions and Future Works}
\label{sec:conclusions}
This study has presented an automated method for labelling source code files through weak labelling. We also achieved multi-granular labelling with a hierarchical aggregation, expanding the file-level labels to both package- and project-level annotations.

We evaluated our approach using a mix of automated metrics, and human evaluation. Results from both assessments have shown that weak labelling is able to capture the application domain of the files effectively. Furthermore, hierarchical aggregation preserves the information captured at the file-level and allows for correctly annotating packages and the project as a whole. Moreover, we have shown how the proposed approach enables the identification of new topics for the projects, not previously included by the originating developers. 

Overall, while being an initial step towards file-level classification, our approach demonstrates promising results and has the potential to be extended and further optimized for more accurate and efficient source code labelling.

As the main future work, we will use these annotations to train ML models that perform classification at all levels. Some approaches include using large language models designed for code to extract features from the files; then, neural networks can be trained to perform the classification. Methods that use the structure can also be used: examples like node classification using graph neural network~\cite{Bronstein2017geometricdeeplearning} would easily apply to our case.

While our study provides valuable insights, several areas could be explored to improve the proposed approach. For example, instead of aggregating the results from the file \revision{to annotate the packages}, one could apply the LF directly to \revision{the package} and then find the best approach to combine \revision{this information, with the annotations of the files belonging to the package.} 

\rev{Moreover, to address the issue of which label refers to core functionalities or side functionality, since we have the dependency graph, we can use community detection algorithms and centrality measures. These metrics can be used to boost topics that are in the more central files.}

Another critical aspect that would make our approach even better is the creation of a hierarchy among the terms in our taxonomy. This will improve the labelling as it can consider relations between labels, reducing cases where a more generic label is preferred given a higher pool of terms due to training data. Again, domain-specific information, like SED-KGraph~\cite{izadi2023sedgraph} and natural language approaches, can be used to achieve this.

\revision{Along with creating a taxonomy, developing a strategy to better assign the keywords extracted from each project to the labels is something to explore better. One approach could use semantic information of the terms to decide to which label the term belongs.}

Furthermore, future work will focus on expanding to more programming languages, like C\# and Python, which have similar project structures to Java projects, expanding the number of projects and increasing the terms list to suit differences between the languages and their communities (\eg Python being the preferred data science, and machine learning programming language). \revision{This extension will allow more data to be used in the training of ML models.}

\revision{Lastly, future work could explore the use of more label-specific LFs, for example, the use of specific LM, or knowledge bases to create even more LFs that target subsets of labels. This in combination with better ensemble approaches could improve the annotation of files.}

\section{Data Availability Statement}
The dataset used and generated artefacts are available in a \textit{Zenodo} repository: \href{https://zenodo.org/record/7943882}{https://zenodo.org/record/7943882}. The code is available at the following repository:  \href{https://github.com/SasCezar/CodeGraphClassification}{https://github.com/SasCezar/CodeGraphClassification}

\section{Conflict of Interest}
The authors declared that they have no conflict of interest.

\bibliography{references}

\end{document}